\documentclass[twocolumn,showpacs,preprintnumbers,amsmath,amssymb,floatfix,prd]{revtex4}
\usepackage{epsfig}
\usepackage{dcolumn}
\begin{document}
%
%\preprint{XYZ}
%
\title{Extraction of Spin-Dependent Parton Densities and
Their Uncertainties}
\author{Daniel de Florian}\email{deflo@df.uba.ar}
\author{Rodolfo Sassot}\email{sassot@df.uba.ar}
\affiliation{Departamento de Fisica, Universidad de Buenos Aires,
Ciudad Universitaria, Pabellon 1 (1428) Buenos Aires, Argentina}
\author{Marco Stratmann}\email{marco@ribf.riken.jp}
\affiliation{Institut f\"{u}r Theoretische Physik,
Universit\"{a}t Regensburg, 93040 Regensburg, Germany \\
Institut f\"{u}r Theoretische Physik und Astrophysik, Universit\"{a}t W\"{u}rzburg,
97074 W\"{u}rzburg, Germany}
\author{Werner Vogelsang}\email{vogelsan@quark.phy.bnl.gov}
\affiliation{Physics Department, Brookhaven National Laboratory, Upton, NY 11973}

%%%%%%%%%%%%%%%%%%%%%%%%
\begin{abstract}
%%%%%%%%%%%%%%%%%%%%%%%%
%
We discuss techniques and results for the extraction of the nucleon's 
spin-dependent parton distributions and their uncertainties from
data for polarized deep-inelastic lepton-nucleon and proton-proton scattering
by means of a global QCD analysis. Computational methods are described that 
significantly increase the speed of the required calculations to a level that 
allows to perform the full analysis consistently at next-to-leading order accuracy.
We examine how the various data sets help to constrain different aspects of the 
quark, anti-quark, and gluon helicity distributions. Uncertainty estimates are performed 
using both the Lagrange multiplier and the Hessian approaches. We 
use the extracted parton distribution functions and their estimated uncertainties
to predict spin asymmetries for high-transverse momentum pion and jet production 
in polarized proton-proton collisions at 500~GeV center-of-mass system energy at BNL-RHIC,
as well as for $W$ boson production.
\end{abstract}
\pacs{13.88.+e, 12.38.Bx, 13.60.Hb, 13.85.Ni}
\maketitle
%
%%%%%%%%%%%%%%%%%%%%%%%%%%%%%%%%%%%%%%%%%%%%%%%%%%%%%%%%%%%%%%%%%%%%%%%%%%
%
%%%%%%%%%%%%%%%%%%%%%%%%
\section{Introduction}
%%%%%%%%%%%%%%%%%%%%%%%%
%
The last twenty years have witnessed remarkable improvements in the
sophistication and precision of methods devoted to the extraction of
parton density and fragmentation functions from experimental hard scattering data.
These distributions are essential ingredients for any phenomenological study
of hard scattering processes involving identified hadrons in the initial and
final-state, respectively. 
Their precise knowledge is not only critical for testing the very successful
framework of perturbative Quantum Chromodynamics (pQCD) but, in more general terms,
also for quantifying uncertainties for precision studies of the Standard Model
and searches of ``new physics'' at high energy accelerators like the CERN-LHC.
At the same time, parton densities and fragmentation functions give fundamental
insights into nucleon structure and the hadronization mechanism.

With the gain in precision and refinements of analyses,
modern parton distribution functions (PDFs) have often revealed
intriguing aspects of hadronic structure, such as
the sizable breaking of isospin symmetry in the light sea quark sector,
suggestions of differences between the strange quark and anti-quark distributions,
the steep rise of the distributions at small momentum fractions,
and an interesting pattern of modifications of the distributions in nuclei, to name
just a few. Certainly one of the most striking results is the unexpectedly small fraction, 
about a quarter, of the proton's spin that can be attributed to the intrinsic 
angular momenta of quarks and anti-quarks. This finding, famously dubbed ``proton 
spin crisis'', has triggered a flurry of experimental and theoretical activity
aiming at clarifying the contributions of gluons and orbital angular momenta 
of partons to the spin of the proton \cite{ref:review}.

The only way to effectively deconvolute the experimental information on PDFs,
which in its raw form is smeared over the light-cone momentum fraction $x$,
summed over many different partonic subprocesses, and taken at different hard scales 
$Q$ for each data point, is a ``global QCD analysis''. It treats all available 
probes simultaneously, in order to extract the set of universal PDFs that yields the 
optimal theoretical description of the combined data. For the case of polarized PDFs, 
the available world-data are from polarized deep-inelastic scattering 
(DIS)~\cite{ref:emc-a1p,ref:smc-a1pd,ref:compass-a1d,ref:e142-a1n,ref:e143-g1pd,
ref:e154-a1n,ref:e155-g1p,ref:e155-g1d,ref:hermes-a1he3-sidisn,ref:hermes-a1pd,
ref:halla-g1n,ref:clas-g1pd}, 
semi-inclusive DIS (SIDIS)~\cite{ref:smc-sidis,ref:hermes-sidispd,ref:hermes-a1he3-sidisn,
ref:compass-sidisd},
photo- and electroproduction of hadrons and charm 
\cite{ref:e155-hadron,ref:hermes,ref:smc,ref:compass-2had,ref:compass-charm},
and proton-proton ($pp$) collisions at BNL-RHIC~\cite{ref:phenix-pion-run5,
ref:phenix-pion-run6,ref:phenix-pion-run6-62gev,ref:star-jets-run5,
ref:star-jets-run6-prel,rhic:charged}.
The different data sets are complementary in the sense that they probe different
aspects of the helicity dependent PDFs. Fully inclusive DIS data
from the many different experiments are pivotal in precisely determining the
sums of quark and anti-quark distributions, SIDIS data help to tell different quark flavors
and quark and anti-quarks apart, and RHIC $pp$ data give a first direct constraint on
gluon polarization.

A global QCD analysis of nucleon spin structure at full next-to-leading (NLO)
accuracy was completed recently \cite{ref:dssv}.
The present paper gives in large part a more detailed account of the methods and 
results of \cite{ref:dssv}. It also addresses the issue of the uncertainties 
of the PDFs in a more detailed and comprehensive way. 
As customary for recent unpolarized PDF analyses \cite{ref:cteq-latest,ref:mstw}, 
we provide sets of polarized PDFs associated with displacements in the PDF parameter space
in the vicinity of the best fit which greatly facilitate the propagation
of PDF uncertainties to any observable of interest.

As a new feature over all previous fits based only on DIS~\cite{grsv,bb,lss,aac,soffer}, 
or combined DIS and SIDIS data \cite{dns}, the analysis \cite{ref:dssv} included 
for the first time also results from polarized $pp$ scattering at RHIC in a NLO 
framework. It benefitted significantly from an 
improved knowledge of parton-to-hadron fragmentation functions \cite{ref:dss} 
which are an essential non-perturbative input for the theoretical description of
all processes with identified hadrons in the final state, such as SIDIS. 
For the first time, these fragmentation functions provide a
good description of identified hadron yields in the entire kinematic 
regime relevant for the analysis of polarized SIDIS and $pp$ data \cite{ref:dss}.
For the time being, hadron and charm production data from fixed-target experiments
\cite{ref:e155-hadron,ref:hermes,ref:smc,ref:compass-2had,ref:compass-charm},
which constrain the gluon polarization at momentum fractions around $x\simeq 0.1$,
were not included in the analysis \cite{ref:dssv} since 
NLO calculations of the relevant cross sections are not yet complete \cite{ref:photoprod}.
We will provide a comparison to these data in order to demonstrate their
consistency with the results of the global fit.
RHIC data for charged pion spin asymmetries \cite{rhic:charged} are also 
not taken into account as they are still preliminary and statistically not
as significant as the neutral pion or jet data \cite{ref:phenix-pion-run5,
ref:phenix-pion-run6,ref:phenix-pion-run6-62gev,ref:star-jets-run5,
ref:star-jets-run6-prel}. 
With sufficient statistics, however, they can provide an important constraint 
on the gluon polarization as will be shown below.

The use of parton distributions in predictions for Standard Model 
benchmark processes, e.g., as ``luminosity candles'' at the LHC,
or in understanding fundamental properties of a nucleon like its spin,
not only requires a careful extraction of PDFs from data but also a proper
assessment of their uncertainties and how they propagate to other observables of interest.
In spite of a great deal of activity and many significant achievements for both
parton distribution and fragmentation functions (FFs), this has shown to be a
rather formidable task in 
practice~\cite{ref:cteq-latest,ref:mstw,ref:unpolpdf,nnpdf,ref:dss}.
The specific challenge of a global QCD analysis is to incorporate a large body of data
from many experiments with diverse characteristics and errors. The complications
are compounded by uncertainties inherent to the theoretical framework used to
describe the data, which are notoriously difficult to quantify. Examples are
the choice of the factorization scale, the functional form
used to parameterize the PDFs, or
unavoidable approximations and assumptions limiting the parameter space.

Several complementary strategies have been devised and
implemented to estimate uncertainties of PDFs and FFs~\cite{cteq0,cteq1,cteq2}.
In general, one starts with introducing an effective $\chi^2$ function
that combines all phenomenological inputs to the analysis as a quantitative measure
of the goodness of the global fit. Minimizing this  $\chi^2$ function yields the optimal
set or ``best fit'' of parameters in the multi-dimensional space defining the PDFs.
The most common method to determine the range of uncertainties
is to study the dependence of $\chi^2$ near its global minimum based on a
Taylor expansion and keeping only the leading term
as characterized by the error matrix or its inverse, the Hessian matrix~\cite{cteq2}.
This {\em{assumes}} a quadratic form in the displacements of all parameters
from their optimum values. The Hessian also determines the uncertainties of any
other physical observable ${\cal{O}}$, provided that the dependence of ${\cal{O}}$
on the fit parameters is approximately linear around the minimum.
Both assumptions are not necessarily adequate in the complex global analysis environment,
and their range of applicability needs to be carefully scrutinized.

The more robust method of Lagrange multipliers~\cite{cteq1} 
circumvents all these shortcomings
and is free of assumptions concerning the 
functional dependence of $\chi^2$ on the fit parameters. 
The idea is to explore directly how the fit to data deteriorates if one enforces
certain values for an observable ${\cal{O}}$ away from its best fit value.
In practice, one performs a series of constrained fits in which $\chi^2$ is minimized for
particular values of ${\cal{O}}$, in order 
to map out the parametric relationship between $\chi^2$
and ${\cal{O}}$. The method is straightforward to implement and can be applied to any
combination of physical observables or even to fit parameters themselves.
We will pursue and compare both methods, Hessian and Lagrange multiplier, 
to estimate uncertainties for the
shape and truncated first moments of helicity-dependent PDFs
using the analysis presented in \cite{ref:dssv} as the starting point.
The Lagrange multiplier technique will provide
the necessary benchmarks for testing the accuracy of
approximations within the Hessian method.
We note that alternative approaches recently proposed in the literature
include studies of uncertainties based on neural networks
or large samples of PDFs generated with Monte-Carlo methods~\cite{nnpdf}.

In any case, all of these methods require an extensive
number of calculations and minimizations of the effective $\chi^2$ function,
in order to explore the very complex and entangled sensitivity of the data
to variations of the parameters describing the PDFs.
This calls for new and more efficient calculational tools to include
all observables used in the global fit consistently at NLO accuracy
without resorting to potentially unreliable approximations.
In particular, numerical computations of NLO cross sections in hadron-hadron
scattering are prone to being very time consuming. 

In order to deal with this problem, Ref.~\cite{ref:dssv} employed a method based 
on the Mellin transform technique proposed in Refs.~\cite{sv,mellinold}, which allows
to speed up the relevant NLO computations to a level that they can be incorporated 
in the global analysis. An important improvement of this Mellin transform method
was the implementation of a novel Monte-Carlo sampling technique. This new computational 
strategy proved already very useful for the analysis of single-inclusive observables in $pp$ 
scattering that are presently relevant at RHIC. However, it is completely general and
becomes especially powerful when less inclusive observables like two-particle correlations
in hadronic collisions need to be incorporated in the global QCD fits, as will soon be
the case. In the present paper we will describe the Mellin transform technique and 
its improvement in detail. 
We note, that the fast and efficient Mellin technique for incorporating NLO $pp$
processes is, of course, not restricted to analyzing helicity-dependent PDFs, but
could equally find important applications for QCD processes at the LHC.

The paper is organized as follows: in the next Section we describe all technical
details of a global PDF analysis. We first discuss the $\chi^2$ function and the underlying 
ideas of the Hessian and Lagrange multiplier approaches for estimating PDF uncertainties.
We next describe in detail our Mellin moment and Monte-Carlo sampling techniques as
implemented for fast evaluations of NLO $pp$ cross sections in our global QCD analysis.
In Sec.~III we apply all techniques to the global analysis of
helicity-dependent PDFs \cite{ref:dssv}. We discuss the results
for the best fit and its uncertainties. We argue that the range of applicability of the 
Hessian method is limited to estimating uncertainties of helicity-dependent PDFs consistent 
with only small departures from the best global fit, 
corresponding to $\Delta \chi^2 \approx 1$.
We present sets of polarized PDFs associated with displacements 
along the eigenvector directions of the Hessian matrix and 
resulting in $\Delta \chi^2 = 1$, which characterize the PDF parameter
space in the vicinity of the global minimum in a process-independent way. We also 
explore the impact of the individual data sets on the results and uncertainties
obtained for helicity-dependent PDFs in the global fit.
In Sec.~IV we study the potential of upcoming measurements at
RHIC at $\sqrt{S}=500$~GeV center-of-mass system (c.m.s.) energy
for further constraining the polarized PDFs. We focus on predictions 
for single-inclusive pion and jet production, and on $W$ 
boson single-spin asymmetries. We conclude in Sec.~V.

%%%%%%%%%%%%%%%%%%%%%%%%%%%%%%%%%%%%%%%%%%%%%%%%%%%%%%%%%%%%%%%%%%
\section{\label{sec:comp}Techniques for NLO global PDF analyses}
%%%%%%%%%%%%%%%%%%%%%%%%%%%%%%%%%%%%%%%%%%%%%%%%%%%%%%%%%%%%%%%%%
%
In this Section, we will describe all techniques we use for the global
analysis of polarized PDFs. The first two Subsections discuss the $\chi^2$ function 
and the various methods for the analysis of PDF uncertainties. Much of the discussion 
here will follow the pioneering work in 
Refs.~\cite{cteq0,cteq1,cteq2}. We then lay out the details of our Mellin 
moment and Monte-Carlo sampling techniques. 
%
%%%%%%%%%%%%%%%%%%%%%%%%%%%%%%%%%%%%%%%%%
\subsection{The effective $\chi^2$ function}
%%%%%%%%%%%%%%%%%%%%%%%%%%%%%%%%%%%%%%%%%
%
Global QCD extractions of PDFs 
\cite{ref:dssv,ref:cteq-latest,ref:mstw,ref:unpolpdf,cteq6,ref:mrst} 
or FFs \cite{ref:dss}
are implemented around an effective $\chi^2$ function
that quantifies the goodness of the fit to data for a given set of
theoretical parameters $\{a_i\}$, $i=1,\ldots,N_{\mathrm{par}}$ that determines 
the PDFs or FFs at some initial scale $\mu_0$.
The simplest $\chi^2$ function, convenient for the search for optimum PDFs by minimization,
is usually taken as
\begin{equation}
\label{eq:chi2def}
\chi^2(\{a_i\}) =\sum_{n=1}^{N_{\mathrm{exp}}} \sum_{j=1}^{N_{\mathrm{data}}^{(n)}}
\omega_j \left( \frac{D_j-T_j(\{a_i\})}{\delta D_{j}} \right)^2 \ ,
\end{equation}
where $N_{\mathrm{exp}}$ counts the individual experimental data sets
and $N_{\mathrm{data}}^{(n)}$ the corresponding number of
data points in each set.
Each data value $D_j$ is compared to the corresponding theoretical estimate $T_j$,
which depends in general non-linearly on the $N_{\mathrm{par}}$ parameters $\{a_i\}$,
weighted with the estimated uncertainties combined in $\delta D_{j}$.
In Eq.~(\ref{eq:chi2def}) $\omega_j$ is a special weighting factor for each data point
with default value one. It can be set to zero if a certain data point is to be removed from
the analysis due to some physics considerations. For instance,
such cuts are routinely introduced in a global fit to remove kinematical regions where the
framework of perturbative QCD used to compute $T_j(\{a_i\})$ is known to be not adequate
for describing the available data.
The simple form (\ref{eq:chi2def}) for $\chi^2$ is appropriate only in the ideal case of data
sets with uncorrelated errors, and $\delta D_j^2$ is then given by 
statistical and point-to-point systematic errors added in quadrature.

For most experiments, additional information on the fully correlated
normalization uncertainty $\delta {\cal{N}}_n$ can be found, i.e.,
on a systematic shift common to the entire data set.
Equation~(\ref{eq:chi2def}) is straightforwardly extended to account for such
normalization uncertainties:
\begin{eqnarray}
\label{eq:chi2defnorm}
\chi^2(\{a_i\}) &=&\sum_{n=1}^{N_{\mathrm{exp}}} \Bigg[ \left(\frac{1-{\cal{N}}_n}{
\delta {\cal{N}}_n}\right)^2 \nonumber \\
&+& \sum_{j=1}^{N_{\mathrm{data}}^{(n)}}
\omega_j \left( \frac{ {\cal{N}}_n D_j-T_j(\{a_i\})}{\delta D_{j}} \right)^2 \Bigg]\,.
\end{eqnarray}
Here, ${\cal{N}}_n$ are the normalization factors,
which can be either fitted along with the $\{a_i\}$,
or even determined analytically in each step of the minimization
by solving $\partial \chi^2 / \partial {\cal{N}}_n = 0$~\cite{cteq1}.

There are several equivalent methods of extending further the simple $\chi^2$ function
in Eq.~(\ref{eq:chi2def}) in the presence of $K^{(n)}$ sources of 
correlated systematic errors for a data point $D_j$ in data set $n$ \cite{cteq1,cteq6}.
The numerically most efficient method treats the correlated systematic errors analytically
in the optimization procedure like for the global normalization
uncertainties discussed in (\ref{eq:chi2defnorm}).
This avoids the construction and inversion of large covariance matrices used in the
conventional approach.
The resulting $\chi^2$ function has the form (assuming $\omega_j=1$ 
for simplicity)~\cite{cteq1,cteq6}
\begin{eqnarray}
\label{eq:chi2defcor}
\chi^2(\{a_i\}) &=&\sum_{n=1}^{N_{\mathrm{exp}}} \Bigg[  \sum_{j=1}^{N_{\mathrm{data}}^{(n)}}
\left( \frac{D_j-T_j(\{a_i\})}{\delta D_{j}^{(u)}} \right)^2
\nonumber \\
&-& \sum_{k,k^{\prime}=1}^{K^{(n)}} B_k \left( A^{-1} \right)_{kk^{\prime}} B_{k^{\prime}}\Bigg]\,,
\end{eqnarray}
where
\begin{eqnarray}
\label{eq:chi2defcoraux1}
B_k &=&  \sum_{j=1}^{N_{\mathrm{data}}^{(n)}}
\frac{ \beta_{kj} (D_j-T_j(\{a_i\})}{\delta D_{j}^{(u)}} \\
\label{eq:chi2defcoraux2}
A_{kk^{\prime}} &=& \delta_{kk^{\prime}} + \sum_{j=1}^{N_{\mathrm{data}}^{(n)}}
\frac{\beta_{kj}\beta_{k^{\prime}j}}{\delta D_{j}^{(u)}}\,.
\end{eqnarray}
Here, $(\delta D_{j}^{(u)})^2$ is the quadratic sum of the 
statistical and uncorrelated systematic errors, and
$\beta_{kj}$ specifies the $k$th correlated
systematic error of data point $D_j$.

We note that most global analyses of unpolarized PDFs now start to include
correlated systematic errors whenever this information is available from
experiment. This is of much importance, as very precise PDF uncertainty
studies are a key ingredient for reliably estimating new physics signals and
standard model backgrounds at the Tevatron or the LHC.
For the time being, and keeping in mind that
our analysis is the first fully global one of this kind for polarized PDFs, we stick to
an effective $\chi^2$ function based on the simplest expression in
Eq.~(\ref{eq:chi2def}). Also, most of the relevant experiments do not
publish the full information on correlated systematic errors.
Whenever a global normalization uncertainty is provided, we have explored the
possibility of normalization shifts to improve the global fit by minimizing
$\chi^2$ according to Eq.~(\ref{eq:chi2defnorm}). We have not found any significant
improvements of the fit from this.
Since data sets are continuously evolving and more and more precise information
becomes available, the proper inclusion of correlated systematic errors
is certainly needed in future global analyses of helicity PDFs.

As mentioned in the Introduction, there are considerable complications when
statistical methods are applied to global QCD analyses based on a
large body of diverse data and a theoretical model with many parameters $\{a_i\}$.
In particular, the statistical value of the definitions given in
Eqs.~(\ref{eq:chi2def})-(\ref{eq:chi2defcor}) has been under
considerable debate~\cite{ref:cteq-latest,ref:mstw,cteq0,cteq1,cteq2,cteq6,ref:mrst}, 
since both the theoretical 
[$T_j(\{a_i\})$] and the experimental inputs [$D_j$, $\delta D_j, \ldots$] 
are far from being ideally suited for statistical analysis.
For instance, uncertainties inherent to the theoretical framework
used to describe the data are notoriously difficult to quantify and usually
correlated. In addition, it is often the case that even in a ``best fit'' to
data, the values of $\chi^2$ per data point for individual experiments
vary considerably around unity, sometimes by much more than the expected amount
$\sqrt{2/N^{(n)}_{\mathrm{data}}}$. Therefore, conclusions drawn on a tolerable 
range of uncertainty from a certain increase $\Delta \chi^2$, must be
considered keeping in mind this complex situation.

%%%%%%%%%%%%%%%%%%%%%%%%%%%%%%%%%%%%%%%%%%%%%%%%%%%
\subsection{Uncertainty estimates: Hessian and Lagrange multiplier methods \label{laghes}}
%%%%%%%%%%%%%%%%%%%%%%%%%%%%%%%%%%%%%%%%%%%%%%%%%%%
%
An important objective is to estimate uncertainties
of the spin-dependent PDFs obtained from the $\chi^2$ optimization.
To this end one must study the behavior of the effective $\chi^2(\{a_i\})$ function
in the neighborhood of the global minimum $\chi^2_0\equiv \chi^2(\{a_i^0\})$,
using reliable statistical methods rather than some
subjective tuning of selected parameters.
Basically two complementary tools have been put forward,
refined to deal with the complexity of global PDF analyses,
and applied to characterize uncertainties in the extraction of
unpolarized PDFs.
We pursue and compare both, the Hessian~\cite{cteq2} and the 
Lagrange multiplier~\cite{cteq1}
methods, in our uncertainty estimates.
We only briefly recall the main elements of the two approaches
and highlight potential problems and limitations.
For a detailed account we refer the reader to Refs.~\cite{cteq0,cteq1,cteq2}.

In the more standard Hessian approach, the exploration of the uncertainties
associated with the fit is implemented through a Taylor expansion 
of $\chi^2(\{a_i\})$ around the global minimum $\chi^2_0(\{a_i^0\})$.
Keeping only the leading quadratic terms, the increase $\Delta \chi^2$
can be written in terms of the Hessian matrix
\begin{equation}
\label{eq:hij}
H_{ij} \equiv \frac{1}{2} \frac{\partial^2 \chi^2}{\partial y_i \partial y_j}
\Bigg|_{0} 
\end{equation}
as
\begin{equation}
\label{eq:chi2hessian}
\Delta \chi^2 =  \chi^2(\{a_i\}) - \chi^2_0(\{a_i^0\}) = 
\sum_{ij} H_{ij} y_i y_j
\end{equation}
where $y_i \equiv  a_i - a_i^0$ and the derivatives in
Eq.~(\ref{eq:hij}) are taken at the minimum.

Global QCD fits are usually characterized by very disparate
uncertainties in different directions of the multi-parameter space,
so that standard methods to evaluate $H_{ij}$ by finite difference, 
as implemented in, e.g., the {\sc Minuit} package~\cite{minuit}, tend to be 
numerically unstable and hence unreliable. 
To overcome such difficulties, a new iterative algorithm
to compute the derivatives in (\ref{eq:hij}) 
was devised in Ref.~\cite{cteq0} and subsequently used in global analyses 
of unpolarized PDFs. We apply this improved Hessian method also in our studies.

Instead of working in the parameter basis $\{a_i\}$, the Hessian 
$H_{ij}$ is expressed in terms of its $N_{\mathrm{par}}$ 
eigenvectors $v_i^{(k)}$ and eigenvalues $\varepsilon_k$.
The displacements $y_i$ in Eqs.~(\ref{eq:hij}) and 
(\ref{eq:chi2hessian}) are replaced by a new set of 
parameters $\{z_i\}$ defined by~\cite{cteq0,cteq2}
\begin{equation}
\label{eq:zi-def}
y_i \equiv \sum_j v_i^{(j)} s_j z_j\,.
\end{equation} 
The $\{z_i\}$ are appropriately normalized by scale factors
$s_j\propto \sqrt{1/\varepsilon_j}$ such that
surfaces of constant $\chi^2$ turn into hyper-spheres in $\{z_i\}$
space, with the distance from the minimum given by
\begin{equation}
\label{eq:deltachi2z}
\Delta \chi^2=  \sum_i z_i^2\,.
\end{equation}
Large eigenvalues $\varepsilon_k$ correspond to steep directions in which $\chi^2$ 
changes rapidly and the parameters are tightly constrained by the data,
while small eigenvalues belong to directions where the parameters are only
weakly determined.

Within the eigenvector representation, it is convenient to construct 2$N_{\mathrm{par}}$ 
eigenvector basis sets of PDFs which greatly facilitate the propagation of PDF
uncertainties to arbitrary observables ${\cal{O}}$~\cite{cteq2}.
These basis sets $S_k^{\pm}$ are defined in $\{z_i\}$ space by
\begin{equation}
\label{eq:eigenset-def}
z_i(S_k^{\pm}) = \pm T \delta_{ik}
\end{equation}
and hence correspond to positive and negative displacements 
along each of the eigenvector directions by the amount $T=\sqrt{\Delta \chi^2}$
still tolerated for an acceptable global fit.
To estimate the error $\Delta{\cal{O}}$ on a quantity ${\cal{O}}$ away from its 
best fit estimate ${\cal{O}}(S^0)$ it is only necessary to evaluate 
${\cal{O}}$ for each of the 2$N_{\mathrm{par}}$ sets $S_k^{\pm}$~\cite{cteq2}, i.e.,
\begin{equation}
\label{eq:obserror-hessian}
\Delta{\cal{O}} = \frac{1}{2} \left[ \sum_{k=1}^{N_{\mathrm{par}}}
[{\cal{O}}(S^+_k) - {\cal{O}}(S^-_k)]^2 \right]^{1/2}\;.
\end{equation}
%
%%%MS Alternatively, one can compute asymmetric bounds by
%
%%\begin{equation}
%%\label{eq:obserror2-hessian}
%%\Delta{\cal{O}}^{\pm} = \frac{1}{2} \left[ \sum_{k=1}^{N_{\mathrm{par}}}
%%[{\cal{O}}(S^{\pm}_k) - {\cal{O}}(S_0)]^2 \right]^{1/2}\;.
%%\end{equation}
%

One must keep in mind that the propagation of PDF uncertainties in
the Hessian method has been derived under the assumption that a 
first order, linear approximation is adequate. Of course, due to the
complicated nature of a global fit, deviations, also from the
simple quadratic behavior in Eq.~(\ref{eq:chi2hessian}),
are inevitable, and error estimates based on the Hessian method 
are not necessarily always accurate.

A strategy based on Lagrange multipliers avoids all these assumptions
and probes the uncertainties on any quantity ${\cal{O}}(\{a_i\})$ 
much more directly. The result is a parametric relationship between
one or more observables ${\cal{O}}_j (\{a_i\})$ and the effective
$\chi^2$ function used to determine the goodness of the global fit.
Its application to global QCD analyses was
proposed in Ref.~\cite{cteq1} not only to estimate uncertainties of
observables depending on PDFs but also to test the range of
applicability of the Hessian approach described above.

In practice, the method is implemented by minimizing a function
\begin{equation}
\label{eq:lagrange}
\Psi(\{a_i\},\{\lambda_j\}) = \chi^2(\{a_i\})+ \sum_j \lambda_j {\cal{O}}_j (\{a_i\})
\end{equation}
with respect to the set of PDF parameters $\{a_i\}$ for fixed values of the 
Lagrange multipliers $\{\lambda_j\}$. Each multiplier is related to one 
specific observable ${\cal{O}}_j$, and the choice $\lambda_j=0$ 
corresponds to the best fit $S^0$.
By repeating this minimization procedure for many values of $\lambda_j$
one can map out precisely how the fit to data
deteriorates as the expectation for the observable ${\cal{O}}_j$
is forced to change. Here, contrary to the Hessian method, no assumptions 
are made regarding the dependence of $\chi^2$ or the observable ${\cal{O}}_j$
on the parameters $\{a_i\}$ of the fit.
The Lagrange multiplier method also generates a large set of
sample PDFs along the curve of maximum variation of the
observable(s) used in Eq.~(\ref{eq:lagrange}).

In principle, the Lagrange multiplier method is superior to the Hessian approach
for reliably estimating uncertainties. For a given
$\Delta \chi^2$ it finds the largest range of ${\cal{O}}_j(\{a_i\})$
allowed by the data used in the global fit and the theoretical ansatz,
independent of the approximations involved in the Hessian method.
The entire parameter space $\{a_i\}$ is explored in minimizing Eq.~(\ref{eq:lagrange}),
not necessarily limited to the vicinity of the best fit $\{a_i^0\}$.
In practice, a large number of global fits is required to map
out all $\chi^2$ profiles of interest. Thanks to our novel Monte Carlo
sampling technique to be described below, this is no longer a
serious limitation computationally. The only drawback to this method is that 
it requires access to the full machinery of global fitting to estimate 
uncertainties of a given observable of interest, contrary to the Hessian method 
for which the eigenvector PDF sets $S_k^{\pm}$ make it very simple to propagate 
PDF uncertainties to arbitrary observables.

In this paper, Lagrange multipliers will mainly be the 
method of our choice for estimating uncertainties 
and to monitor to what extent the approximations involved in the 
Hessian approach are justified.

%%%%%%%%%%%%%%%%%%%%%%%%%%%%%%%%%%%%%%%%%%%%%%%%%%%
\subsection{Computational technique: The Mellin method}
%%%%%%%%%%%%%%%%%%%%%%%%%%%%%%%%%%%%%%%%%%%%%%%%%%%
%
As is evident from the previous Subsections, 
the extraction of PDFs in a global QCD analysis of a large body of data
at NLO accuracy requires an extensive number of time consuming computations
of the corresponding observables in each step of the $\chi^2$ minimization procedure.
The large number of parameters specifying the functional form of the PDFs in the fit,
typically of ${\cal{O}}(20)$, and the need for 
a proper assessment of their uncertainties, add to this.
Assuming a representative analysis with about 500 data points, 5000 iterations
to reach the optimum set of parameters, and a modest computing time of 1 second per
cross section calculation at NLO, one easily realizes that computational improvements
are very much in demand.

We stress at this point that approximating NLO corrections by parameterizing them by 
a $K$-factor, $K \equiv \sigma^{\mathrm{NLO}}/\sigma^{\mathrm{LO}}$, which is a
possible strategy in the spin-averaged case in order to speed up the 
analysis~\cite{ref:cteq-latest,ref:mstw,ref:unpolpdf,nnpdf,cteq6,ref:mrst},
is not a viable option in the polarized case. In the latter, the parton distributions
as well as the hard scattering cross sections may have nodes and change sign within
the kinematic region of interest. As a result, different partonic subprocess contributions 
can have very different NLO corrections that are never well approximated by a single 
$K$-factor. This problem is even more prominent at the present time when the 
spin-dependent parton distributions, in particular the gluon distribution, are still 
poorly known. Also, from a more theoretical point of view, the NLO corrections are
expected to decrease the factorization/renormalization scale dependence of the 
calculation. In the ideal case that the NLO cross section has relatively little scale
dependence, this would imply that the $K$-factor inherits the full LO scale dependence 
and thus cannot serve well as a proxy for the NLO corrections. We thus do not really
see a useful workaround that would avoid inclusion of the full NLO calculation in the 
global analysis.

The required numerical calculations also involve the scale evolution of the
PDFs from some initial scale $\mu_0\sim {\cal{O}}(1\,\mathrm{GeV})$ to each of the
scales relevant for the data points.
The evolution is governed by a well-known set of integro-differential 
equations~\cite{ap,ap1} that can be solved analytically after an integral 
transform from Bjorken $x$ space to complex Mellin $N$-moment space.
The Mellin transform of a generic function $\varphi$ depending on the light-cone momentum
fraction $x$ is defined as
\begin{equation}
\label{eq:mellin}
\Phi(N) \equiv \int_0^1 x^{N-1} \varphi(x)\, dx\,,
\end{equation}
and its inverse reads
\begin{equation}
\label{eq:invmellin}
\varphi(x) \equiv \frac{1}{2\pi i} \int_{{\cal{C}}_N} x^{-N} \Phi(N)\, dN\,.
\end{equation}
Here ${\cal{C}}_N$ denotes a suitable contour in the complex $N$ plane
that has an imaginary part ranging from $-\infty$ to $+\infty$ and that intersects the real
axis to the right of the rightmost pole of $\Phi(N)$. In practice, it is
beneficial to choose the contour to be bent at an angle $<\pi/2$ towards the
negative real-$N$ axis~\cite{angle}. The integration in (\ref{eq:invmellin}) 
can then be very efficiently performed numerically by choosing the values of 
$N$ as the supports for a Gaussian integration.

The transformation (\ref{eq:mellin}) has the welcome property that convolutions
factorize into ordinary products, which greatly simplifies calculations based on
Mellin moments. It can be carried out analytically not only for the splitting functions
governing the scale evolution of the PDFs but even for the
partonic hard scattering cross sections
for both inclusive and semi-inclusive DIS. The latter case requires straightforward
extensions of Eqs.~(\ref{eq:mellin}) and (\ref{eq:invmellin}) to double transformations
as was discussed in Ref.~\cite{ref:aemp,sv}.
This reflects the fact that SIDIS depends on two scaling variables, the momentum fractions
$x$ and $z$ taken by the struck parton from the parent nucleon and by the final-state hadron 
from the scattered parton, respectively.
The usefulness of the Mellin technique was demonstrated in practice in a global analysis
of helicity-dependent PDFs using all available DIS and SIDIS data~\cite{dns}.

However, the inclusion of observables in hadron-hadron collisions or in less inclusive 
reactions in lepton-hadron scattering, which are crucial for determining, e.g., the 
gluon distribution, require a more elaborate framework. They involve multiple 
convolution integrals between one or more PDFs, partonic cross sections, and, depending 
on the process, fragmentation functions. More importantly, they typically depend on 
several kinematic variables, so that there is no direct way of taking 
Mellin moments of the cross section under which it would become a simple product of 
Mellin moments of PDFs and partonic cross sections. An example is single-inclusive pion 
production in proton-proton collisions, $pp\to
\pi X$, at measured transverse momentum $p_T$ and rapidity $\eta$ of the pion. While
$x_T=2 p_T/\sqrt{S}$ is the typical scaling variable of the process, taking Mellin
moments of the cross section in $x_T^2$ does not lead to a simple expression involving
the moments of the PDFs. An efficient computational scheme that allows to circumvent 
this complication has been devised in Ref.~\cite{sv,mellinold}. Its main feature
is to use the inverse Mellin transforms of the PDFs in the theoretical expression,
which makes it possible to store all numerically time consuming calculations involving 
the lengthy and complicated expressions for the underlying hard scattering cross sections 
on large ``look-up tables'' or ``grids'' in Mellin moment space. Here, we review the 
technique of~\cite{sv} and describe a method to compute these grids very efficiently 
using Monte-Carlo sampling techniques. The latter 
allows to generalize the Mellin moment technique
beyond the single inclusive processes considered in Ref.~\cite{sv}.

A typical observable of interest, the spin-dependent cross section for $pp\to
\pi X$ at RHIC, has the following general factorized structure:
\begin{eqnarray}
\label{eq:xsecdef}
\Delta\sigma\big|_{\mathrm{bin}}
&\equiv& \frac{1}{2} \left[ \Delta\sigma\big|_{\mathrm{bin}}(++) -
\Delta\sigma\big|_{\mathrm{bin}}(+-) \right]
\\
\label{eq:skeleton}
&=&
\sum_{i,j,k} \int \Delta f_i(x_1)\, \Delta f_j(x_2) \,D_{k}(z) \nonumber \\
&&\; \times \,  d\Delta\hat{\sigma}_{ij\to kX}(x_1,x_2,z) \, {\cal S}\, dx_1 dx_2 dz\;.
\end{eqnarray}
Here we have suppressed the dependence on kinematic variables other than momentum 
fractions, and on the factorization and renormalization scales.
In (\ref{eq:skeleton}), $\Delta f_i$ and $D_k$ denote the spin-dependent parton
distribution and spin-averaged
fragmentation functions for flavor $i$ and $k$, respectively,
and $d\Delta \hat{\sigma}_{ij\to kX}$ the relevant partonic hard scattering cross sections.
${\cal S}$ represents a ``measurement function'' that defines details of the
observable $\Delta\sigma\left|_{\mathrm{bin}}\right.$ in each bin,
such as the kinematical ranges explored and the relevant experimental cuts.
The symbols $\pm$ in Eq.~(\ref{eq:xsecdef}) denote the helicity combinations
of the colliding longitudinally polarized protons defining the cross section
$\Delta\sigma\left|_{\mathrm{bin}}\right.$.

Following the strategy developed in Ref.~\cite{sv}, we replace the PDFs in 
Eq.~(\ref{eq:skeleton}) by their representations as Mellin inverses, 
Eq.~(\ref{eq:invmellin}), and subsequently interchange integrations to obtain
\begin{eqnarray}
\label{eq:xsecmellin}
\nonumber
\Delta\sigma\big|_{\mathrm{bin}} &=&
-\frac{1}{4\pi^2}\sum_{ijk}
\int_{{\cal C}_M} \int_{{\cal C}_N} L_{ij}(N,M) \\
&& \times d\Delta\hat{\tilde{\sigma}}_{ijk}(N,M)\,dN dM\,,
\end{eqnarray}
where
\begin{equation}
\label{eq:lumipdf}
L_{ij}(N,M) \equiv \Delta f_i(N)\, \Delta f_j(M)
\end{equation}
and
\begin{eqnarray}
\label{eq:xsectilde}
\nonumber
d\Delta\hat{\tilde{\sigma}}_{ijk}(N,M)&\equiv&
\int dx_1 dx_2 dz\, x_1^{-N} x_2^{-M} D_k(z) \\
&& \times\, d\Delta\hat{\sigma}_{ij\to kX}(x_1,x_2,z)\,{\cal S}\,.
\end{eqnarray}
Here and in the following, we assume that only the spin-dependent PDFs
are subject to the global
analysis and that the fragmentation functions $D_k$ are known.

In the next step, one can now calculate the quantities
$d\Delta\hat{\tilde{\sigma}}_{ijk}(N,M)$ in Eq.~(\ref{eq:xsectilde})
{\em prior} to the actual fit, as they do not depend on the PDFs,
and store them in large look-up tables in the Mellin variables $M$ and $N$
along the contours ${\cal{C}}_{M,N}$. In practice, it is convenient to 
choose $M$ and $N$ on the supporting points of a Gaussian integration, see Ref.~\cite{sv} 
for details. As can be seen from~(\ref{eq:xsectilde}), this effectively amounts to 
computing the NLO cross sections for all partonic subprocesses using complex 
``dummy PDFs'' $x_1^{-N} x_2^{-M}$. Even after making use of all symmetry 
relations, e.g., among different subprocesses $ij\rightarrow kX$,
setting up all look-up tables for a typical $N \times M$
grid size of $64\times 64$ is a rather time-consuming step
in practice, although it has to be done only once. 
This is where our new Monte-Carlo sampling technique
has considerable advantages over a ``brute-force'' computation.

First, we recall that a Monte-Carlo algorithm reduces
the multiple integrations in (\ref{eq:xsecdef})
to a finite sum over random ``events'' $n$,
\begin{equation}
\label{eq:mcsum}
\Delta\sigma\big|_{\mathrm{bin}} =
\frac{1}{\kappa}\sum_{n=1}^I \,w^{(n)} {\cal S}
\end{equation}
with weights $w^{(n)}$ and scaled by the number of iterations, $\kappa$,
used to generate large samples of $I$ events. The weight includes the
dependence on the parton distribution and fragmentation functions and
the hard scattering cross sections for each event.
In Eq.~(\ref{eq:mcsum}) we keep the measurement function
${\cal S}$ separate from $w^{(n)}$, as it is usually implemented 
only after an event has been generated.

The most efficient strategy to compute all the required
$d\Delta\hat{\tilde{\sigma}}_{ijk}(N,M)$ in Eq.~(\ref{eq:xsectilde})
is to choose a suitable set of trial PDFs $\Delta f_i$ and
to perform a {\it single} high-statistics Monte-Carlo integration 
calculation of the cross section in Eq.~(\ref{eq:skeleton}). During 
the calculation one stores, for each event $n$, the momentum fractions 
$x_{1,2}^{(n)}$ and the corresponding weights $w^{(n)}_{ijk}$ for all of 
the subprocesses $ij\rightarrow kX$. Renormalizing each weight by 
\begin{equation}
\label{eq:newweight}
w^{(n)\prime}_{ijk} \equiv w^{(n)}_{ijk}/L^{(n)}_{ij}
\end{equation}
with 
\begin{equation}
\label{eq:lumipdfx}
L^{(n)}_{ij} \equiv \Delta f_i(x_1^{(n)})\Delta f_j(x_2^{(n)})\,,
\end{equation}
removes its dependence on the assumed set of PDFs. The 
$d\Delta\hat{\tilde{\sigma}}_{ijk}(N,M)$ are then straightforwardly 
obtained as
\begin{equation}
\label{eq:sigmasum}
d\Delta\hat{\tilde{\sigma}}_{ijk}(N,M) \equiv
\frac{1}{\kappa} \sum_{n=1}^I
\left(x_1^{(n)}\right)^{-M} \left(x_2^{(n)}\right)^{-N} 
\,w^{(n)\prime}_{ijk}\, {\cal S}\,.
\end{equation}
In other words, knowledge of the set $x_1^{(n)},x_2^{(n)},w^{(n)\prime}_{ijk}$,
which corresponds to a profile of the integrand in Eq.~(\ref{eq:xsectilde}),
allows to simultaneously 
compute the moments $d\Delta\hat{\tilde{\sigma}}_{ijk}(N,M)$ for
{\it all} $N,M$, without having to do an individual Monte-Carlo integration
for each of them as in~(\ref{eq:xsectilde}). This greatly reduces the computational 
burden. To give an example, for our global analysis of polarized PDFs, which
currently includes about 50 very time-consuming NLO calculations of
single-inclusive jet and pion production at RHIC
in each step of the $\chi^2$ minimization,
all the necessary grids in Eq.~(\ref{eq:sigmasum})
can be computed within approximately one day of CPU time on a {\em single} 
standard PC with a CoreDuo processor running at 2 GHz. Once these grids are 
available, a full PDF fit can be finalized in about 15--30 minutes.
A detailed PDF uncertainty assessment, which requires a very large number of 
$\chi^2$ minimizations, can then be performed easily in about 1--2 weeks.
We note that in practice one does not even need to physically store the 
set $x_1^{(n)},x_2^{(n)},w^{(n)\prime}_{ijk}$, since the 
summations~(\ref{eq:sigmasum}) can in fact be done simultaneously with the 
generation of the events $n$. 

The formula in Eq.~(\ref{eq:mcsum}) applies to any computation of the theoretical 
cross section when a Monte-Carlo integration is employed. 
This is the case for analytical NLO 
calculations of single-inclusive jet or hadron rates~\cite{anjet,anpion} where Monte-Carlo 
integration techniques are just used to perform the convolution 
with the parton distributions, 
or for NLO parton-level Monte-Carlo generators evaluating general infrared-safe 
observables~\cite{mcjet,mchad}.
Therefore our method based on Mellin moments and Monte-Carlo sampling techniques
outlined in Eqs.~(\ref{eq:mellin})-(\ref{eq:sigmasum}) above is completely general
and can be straightforwardly applied to any observable for which a perturbative QCD 
description is available. It can be applied to global analyses of polarized PDFs, 
pursued in this paper, extractions of fragmentation functions, see Refs.~\cite{ref:dss}, 
but equally well to analyses of ordinary spin-averaged PDFs incorporating Tevatron and 
future LHC data consistently at NLO or beyond.
We note that for the latter case an alternative method, called ``fastNLO''~\cite{fastnlo},
has been developed, which also allows to include NLO corrections to hadronic
scattering in a fast and efficient way in a global PDF analysis. Like our method,
it amounts to preparing huge look-up tables that contain all the time-consuming NLO 
calculations prior to the actual fit. In the ``fastNLO'' case, this is done in 
$x$-space, and interpolations between the various support points in $x$ are used. 
This step appears to be rather time-consuming and the method has so far been tested only
for inclusive jet production data. In addition, the evolution of the PDFs needs
to be performed as a separate calculation, whereas in our approach it is immediately 
included in Mellin moment space. 

This latter advantage can in fact be used for further improvements of our Mellin technique.
The scale dependence of the PDFs, which we have so far suppressed, can be schematically 
written as
\begin{equation}
\label{eq:evomell}
\Delta f_i(N,\mu)=\sum_{i^{\prime}} E_{ii^{\prime}}(N,\mu_0,\mu)\,\Delta f_{i^{\prime}}(N,\mu_0)\,.
\end{equation}
Here, $E_{ii^{\prime}}(N,\mu_0,\mu)$ denotes the appropriate evolution matrix 
from the initial scale $\mu_0$ where we parameterize the PDFs to the scale $\mu$
relevant for a certain observable ${\cal{O}}$.
Inserting (\ref{eq:evomell}) into Eq.~(\ref{eq:xsecmellin}) allows to absorb also 
the scale evolution of the PDFs into the pre-calculated Mellin grids by 
extending Eq.~(\ref{eq:sigmasum}) to
\begin{eqnarray}
\label{eq:sigmascale}
d\Delta\hat{\tilde{\sigma}}_{ijk}(N,M) &\equiv&
\frac{1}{\kappa} \sum_{n=1}^I \sum_{i^{\prime}j^{\prime}} x_1^{-M} x_2^{-N}
\nonumber \\
&&\!\!\!\!\!\!\!\!\!\!\!\!\!\!\!\!\!\!\!
E_{i^{\prime}i}(N,\mu_0,\mu) E_{j^{\prime}j}(M,\mu_0,\mu)
\,w^{(n)\prime}_{i^{\prime}j^{\prime}k}\,{\cal S} \ ,
\end{eqnarray}
so that the luminosity function in Eq.~(\ref{eq:lumipdf}) now only refers
to the PDFs at the initial scale $\mu_0$, $L_{ij}(N,M) = \Delta f_i(N,\mu_0)\, 
\Delta f_j(M,\mu_0)$.
An advantage of this re-shuffling is that now all dependence on the scale $\mu$
is contained in the look-up tables (\ref{eq:sigmascale}), eliminating the
need to perform the scale evolution later in the fitting code.
More importantly, if the experimental observable used in the global fit involves
an integration over a bin in, say, the transverse momentum $p_T$ of a jet,
which also acts as the factorization scale, the $\mu$ dependence
of the PDFs is correctly taken care of in the integration.
While we have not made use of this particular improvement in the
present analysis, we expect it to be useful in the future for further
optimizing the performance of the global analysis code.

%%%%%%%%%%%%%%%%%%%%%%%%%%%%%%%%%%%%%%%%%%%%%%%%%%%%%%%%%%%%%%%%%%%%%%%%%%%%%%%%%%%
\section{Global analysis and uncertainty estimates for polarized PDFs}
%%%%%%%%%%%%%%%%%%%%%%%%%%%%%%%%%%%%%%%%%%%%%%%%%%%%%%%%%%%%%%%%%%%%%%%%%%%%%%%%%%%
%
In this Section we give a detailed account of the first
global analysis of polarized PDFs presented
in Ref.~\cite{ref:dssv} which in the following will be
referred to as ``DSSV''.
We first discuss the data selection and the determination 
of the best fit, which we compare to the fitted data. 
We then focus on the studies of uncertainties, including
a comparison of the Lagrange multiplier method used in \cite{ref:dssv}
and the more standard Hessian error matrix approach.
For the latter we present a new family
of eigenvector PDFs, as described above, which greatly facilitates 
estimates of the PDF uncertainties of any given observable of interest.

%%%%%%%%%%%%%%%%%%%%%%%%%%%%%%%%%%%%%%%%%%
\subsection{Determination of the optimal fit \label{secIIIa}}
%%%%%%%%%%%%%%%%%%%%%%%%%%%%%%%%%%%%%%%%%%
%
%%%%%%%%%
% TABLE I
%%%%%%%%%
\begin{table}[th!]
\caption{\label{tab:chi2table} Data used in our NLO global analysis
of polarized parton densities, the individual $\chi^2$ values for
each set, and the total $\chi^2$ of the fit.}
\begin{ruledtabular}
\begin{tabular}{lccc}
experiment& process  & $N_{\mathrm{data}}$ & $\chi^2$ \\
          &          &      &          \\\hline
%
%DIS
%
EMC \cite{ref:emc-a1p}          & DIS (p) &  10  &  3.9  \\
SMC \cite{ref:smc-a1pd}         & DIS (p) &  12  &  3.4  \\
SMC \cite{ref:smc-a1pd}         & DIS (d) &  12  & 18.4  \\
COMPASS \cite{ref:compass-a1d}  & DIS (d) &  15  &  8.1  \\
E142 \cite{ref:e142-a1n}        & DIS (n) &   8  &  5.6  \\
E143 \cite{ref:e143-g1pd}       & DIS (p) &  28  & 19.3  \\
E143 \cite{ref:e143-g1pd}       & DIS (d) &  28  & 40.8  \\
E154 \cite{ref:e154-a1n}        & DIS (n) &  11  &  4.5  \\
E155 \cite{ref:e155-g1p}        & DIS (p) &  24  & 22.6  \\
E155 \cite{ref:e155-g1d}        & DIS (d) &  24  & 17.1  \\
HERMES \cite{ref:hermes-a1he3-sidisn}   & DIS (He)&   9  &  6.3  \\
HERMES \cite{ref:hermes-a1pd}   & DIS (p) &  15  & 10.5  \\
HERMES \cite{ref:hermes-a1pd}   & DIS (d) &  15  & 16.9  \\
HALL-A \cite{ref:halla-g1n}     & DIS (n) &   3  &  0.2  \\
CLAS \cite{ref:clas-g1pd}       & DIS (p) &  10  &  5.9  \\
CLAS \cite{ref:clas-g1pd}       & DIS (d) &  10  &  2.5  \\ \hline
%
%SIDIS
%
SMC \cite{ref:smc-sidis}       & SIDIS (p,\,$h^{+}$)   &  12  &  18.7   \\
SMC \cite{ref:smc-sidis}       & SIDIS (p,\,$h^{-}$)   &  12  &  10.6   \\
SMC \cite{ref:smc-sidis}       & SIDIS (d,\,$h^{+}$)   &  12  &   7.3   \\
SMC \cite{ref:smc-sidis}       & SIDIS (d,\,$h^{-}$)   &  12  &  14.1   \\
HERMES \cite{ref:hermes-sidispd}   & SIDIS (p,\,$h^{+}$)   &   9  &   6.4   \\
HERMES \cite{ref:hermes-sidispd}   & SIDIS (p,\,$h^{-}$)   &   9  &   4.9   \\
HERMES \cite{ref:hermes-sidispd}   & SIDIS (d,\,$h^{+}$)   &   9  &  11.4   \\
HERMES \cite{ref:hermes-sidispd}   & SIDIS (d,\,$h^{-}$)   &   9  &   4.5   \\
HERMES \cite{ref:hermes-a1he3-sidisn}   & SIDIS (He,\,$h^{+}$)  &   9  &   4.7   \\
HERMES \cite{ref:hermes-a1he3-sidisn}   & SIDIS (He,\,$h^{-}$)  &   9  &   6.9   \\
HERMES \cite{ref:hermes-sidispd}   & SIDIS (p,\,$\pi^{+}$) &   9  &   9.6   \\
HERMES \cite{ref:hermes-sidispd}   & SIDIS (p,\,$\pi^{-}$) &   9  &   4.9   \\
HERMES \cite{ref:hermes-sidispd}   & SIDIS (d,\,$\pi^{+}$) &   9  &   9.4   \\
HERMES \cite{ref:hermes-sidispd}   & SIDIS (d,\,$\pi^{-}$) &   9  &  19.5   \\
HERMES \cite{ref:hermes-sidispd}   & SIDIS (d,\,$K^{+}$)   &   9  &   6.2   \\
HERMES \cite{ref:hermes-sidispd}   & SIDIS (d,\,$K^{-}$)   &   9  &   5.8   \\
HERMES \cite{ref:hermes-sidispd}   & SIDIS (d,\,$K^{+}\!\!+\!K^{-}$)  &   9    &   3.4   \\
COMPASS \cite{ref:compass-sidisd}  & SIDIS (d,\,$h^{+}$)       &  12  &   6.2   \\
COMPASS \cite{ref:compass-sidisd}  & SIDIS (d,\,$h^{-}$)       &  12  &   12.0  \\ \hline
%
%RHIC
%
PHENIX \cite{ref:phenix-pion-run5}       & pp ($200\,\mathrm{GeV}$,\,$\pi^0$) &  10 & 14.2    \\
PHENIX \cite{ref:phenix-pion-run6}       & pp ($200\,\mathrm{GeV}$,\,$\pi^0$) &  10 & 7.1 [13.8]
\footnote{The PHENIX data were still preliminary when the
global analysis \cite{ref:dssv} presented here was performed. 
The $\chi^2$ value quoted in brackets is evaluated
with the published data \cite{ref:phenix-pion-run6,ref:phenix-pion-run6-62gev}
but without re-fitting.} \\
PHENIX \cite{ref:phenix-pion-run6-62gev} & pp ($62\,\mathrm{GeV}$,\, $\pi^0$) &   5 & 3.1 [2.8]$^{\,\,a}$ \\
STAR   \cite{ref:star-jets-run5}         & pp ($200\,\mathrm{GeV}$, jet)   &  10 & 8.8  \\
STAR (prel.) \cite{ref:star-jets-run6-prel} & pp ($200\,\mathrm{GeV}$, jet)   &   9 & 6.9  \\ \hline
%
%SUM
%
{\bf TOTAL:} & & 467 & 392.6  \\
\end{tabular}
\end{ruledtabular}
\end{table}
Our first physics objective is to establish the
set of polarized PDFs that gives the optimum theoretical
description of the available hard scattering data.
The data sets for the spin asymmetries we use in our analysis 
are listed in Tab.~\ref{tab:chi2table}, along with the number of 
data points fitted.
We minimize the effective $\chi^2$ function in Eq.~(\ref{eq:chi2def}).
Attempts to further improve the global fit by introducing
normalization shifts for each experiment and 
minimizing $\chi^2$ according to Eq.~(\ref{eq:chi2defnorm}) were to no avail.
All theoretical spin asymmetries in Eq.~(\ref{eq:chi2def})
are calculated at NLO, using the 
appropriate factorized leading-twist expressions. We use the
$\overline{\mathrm{MS}}$ scheme throughout, and all our results
for the polarized PDFs will refer to this scheme. 

In case of inclusive DIS,
the asymmetries are computed, as in our previous analyses~\cite{grsv,dns}, 
as the ratios between the polarized and unpolarized structure functions,
\begin{equation}
\label{eq:A1}
A_1(x,Q^2)= \frac{g_1 (x,Q^2)}{F_1 (x,Q^2)} \ ,
\end{equation}
with
\begin{eqnarray}
\label{eq:g1}
g_1&=& 
\frac{1}{2}\sum_q e^2_q \left\{ \Delta q +\Delta \bar{q}\right.  \nonumber \\
&+&\left.\frac{\alpha_s}{2\pi} \left[
\Delta C_q \otimes \left( \Delta q +\Delta \bar{q}\right)
+ \Delta C_g \otimes \Delta g \right] \right\} \ ,
\end{eqnarray}
and the corresponding expression for $F_1 (x,Q^2)$, both computed at NLO using
the appropriate coefficient functions~\cite{fp,ap1}. For DIS off a deuteron target,
we take into account the $\omega_D=5.8\%$ $D$-wave state probability in relating the 
$g_1$ structure function of the deuteron to those of proton and neutron:
$g_1^d=(1-1.5 \omega_D)(g_1^p+g_1^n)/2$. The extension of the 
expression in~(\ref{eq:A1}) to SIDIS is straightforward, using the NLO coefficient
functions given in~\cite{lambda}. 

For the case of $pp$ scattering, 
the spin asymmetries are computed using the generic expression in Eq.~(\ref{eq:skeleton}) 
at NLO, and its spin-averaged counterpart. The NLO corrections for high-$p_T$
single-jet and hadron production can be found in~\cite{anjet,anpion}, respectively.
We have always chosen the renormalization and factorization scales as the transverse 
momentum $p_T$ of the observed final state, a choice that leads to good agreement
of NLO calculations \cite{anjet,anpion,ref:dss} and experimental data from RHIC 
\cite{ref:starjet,ref:phenix-pion-run5} in the spin-averaged case.
For the computation of the unpolarized cross sections, we always use the NLO 
unpolarized PDFs of Ref.~\cite{ref:mrst}. Whenever fragmentation functions are
needed, as is the case for SIDIS and the RHIC $pp\to \pi X$ data, 
we use the DSS set~\cite{ref:dss} for pions, kaons, and 
unidentified charged hadrons, which was recently obtained from a
global analysis of hadron production data. The use of up-to-date
fragmentation functions that are consistent with HERMES and RHIC
unpolarized data and have quantified uncertainty estimates~\cite{ref:dss} is a 
crucial ingredient of our analysis. It is a major difference with respect to 
the polarized PDF analysis in Ref.~\cite{dns}, where also SIDIS
data were included. In the computation of the $\chi^2$ contribution from SIDIS
asymmetries we have taken into account the uncertainty coming from 
the set of fragmentation functions. In practice, this was done by determining for
each data point the maximum variation of the theoretical estimates $T_j$ in 
Eq.~(\ref{eq:chi2def}) due to the FFs within their own uncertainty ranges quoted 
in~\cite{ref:dss}. This variation was treated as an additional uncertainty and added
in quadrature to the experimental error $\delta D_{j}$. Since at least for pions 
the FFs are fairly well determined, this amounts to additional uncertainties 
of a few percent at most. We do note, however, that this uncertainty estimate
does not reflect possible systematic problems in interpreting the available 
spin-averaged kaon SIDIS data within a leading-twist pQCD analysis, a point to 
which we will return later. 

We do not systematically include any higher twist (or, more generally, 
power-suppressed) contributions in the theoretical calculations that enter our analysis, 
neither for the inclusive and semi-inclusive DIS observables, nor 
for proton-proton collisions. 
We employ a cut of $Q^2>1\,\mathrm{GeV}^2$ for all DIS and SIDIS data, 
and of $p_T>1\,\mathrm{GeV}$ for the RHIC high-$p_T$ polarized $pp$ data.
These cuts have the purpose to exclude regions where contributions
beyond the leading twist, factorized framework of pQCD become 
crucially important. For example, for the SIDIS and RHIC $pp$ data, it 
is known that the underlying unpolarized cross sections in the same 
kinematic domain, i.e., for scales above 1~GeV, can be quite successfully 
described by pQCD~\cite{ref:dss}. 

That said, it is known~\cite{smc} that the relation 
between $A_1$ and $g_1/F_1$ in Eq.~(\ref{eq:A1}) is corrected by a
factor $(1+\gamma^2)\equiv (1+4 M^2x^2/Q^2)$ on the right-hand-side (r.h.s.), 
corresponding to a target mass correction. It has been pointed out~\cite{lss1} 
that this correction is non-negligible in some kinematic regimes accessed
by the lower energy fixed-target experiments, typically at
relatively low $Q^2$ and high $x$. We have therefore corrected for
this factor where necessary. Specifically, since our set of polarized PDFs 
is defined and related to the measured asymmetries through the leading-twist 
relations (\ref{eq:A1}) and (\ref{eq:g1}), our choice is to multiply
data sets that are given in terms of measured $g_1/F_1$ by the factor
$(1+\gamma^2)$, but to leave data sets for measured $A_1$ unchanged. 
The resulting data are confronted with the NLO leading twist calculation.
We stress that various other choices have been adopted in the 
literature~\cite{bb,lss,aac}, using, for example, parameterizations of experimental
data for $F_2(x,Q^2)$ and $R(x,Q^2)=F_2/(2 xF_1)-1$. 
All choices are equivalent 
in the strict leading twist sense, but will in general differ in the amount of 
power corrections needed to describe the data at lower $Q^2$ and/or higher $x$. 
As we shall see below, we find that for our choice without inclusion of any 
power corrections an excellent description of all data sets within our specified 
cuts is achieved (cf.\ also Tab.~\ref{tab:chi2table}), 
with no visible discrepancies even at rather low scales.
For the time being, we thus regard our choice as one that ``empirically'' 
alleviates the need for power-suppressed corrections in spin asymmetries. 
Despite some significant progress on the analysis of higher twist effects
in polarized DIS~\cite{lss,lss1}, further detailed investigations will 
be needed in this area, including implementation of the full target-mass 
corrections~\cite{tmc}, the effects of yet higher orders~\cite{gr,vm} 
which generally reduce the need for higher twist contributions~\cite{gr,bb1},
and so forth.

In order to find the optimal helicity-dependent PDFs from a $\chi^2$ 
minimization, we parameterize them at an input scale of $\mu_0=1\,\mathrm{GeV}$ 
by the following flexible form~\cite{ref:dssv}
\begin{equation}
\label{eq:pdf-input}
x\Delta f_i(x,\mu^2_0) = N_i x^{\alpha_i} (1-x)^{\beta_i}
(1+\gamma_i \sqrt{x}+\eta_i x)\,,
\end{equation}
with independent parameters for $\Delta u+\Delta \bar{u}$, $\Delta d+\Delta \bar{d}$,
$\Delta \bar{u}$, $\Delta \bar{d}$, $\Delta \bar{s}\equiv\Delta s$, and $\Delta g$
(note that here and in the following we interchangeably use $\Delta f_u=
\Delta u$, $\Delta f_g=\Delta g$ etc.~to denote the polarized PDFs).
The minimization is carried out with respect to the set 
of parameters $\{a_i\}=\{N_i,\alpha_i,\beta_i,\gamma_i,\eta_i\}$. 
The PDFs are evolved to the scales $\mu>\mu_0$ relevant in experiment.
The particular functional form and the value for
$\mu_0$ are not too crucial, as long as the parameterization is
flexible enough to accommodate all hard scattering data within their
ranges of uncertainties.
The ansatz (\ref{eq:pdf-input}) deviates considerably from the standard form
used in fits to DIS data only~\cite{grsv,bb,lss,aac,dns}, 
where $\gamma_i=\eta_i=0$, inasmuch as it allows the PDFs to develop nodes and 
to deviate from an SU(3) flavor symmetric sea. As will be seen from our results 
presented below, this extra freedom in parameter space $\{a_i\}$
is crucial in a comprehensive analysis of DIS, SIDIS, and RHIC $pp$ data.

In addition to the much more flexible input parameterization proposed in the 
preceding paragraph, we have repeated the analysis with alternative 
parameterizations, some of them even more flexible than the one we choose.
For example, we have chosen the powers of $x$ in the last two terms different
from $1/2$ and $1$, even allowing the fit to vary them.
None of these modifications resulted in any significant improvement 
in the quality of the fit to data, or changes of the uncertainty bands. 
This indicates that the present data is not really able to discriminate 
between various forms of the input distributions, as long as a sufficiently
flexible choice is made. Therefore our present choice does not introduce 
large additional uncertainty in that respect. 

Analyses of polarized PDFs routinely use constraints that can be derived 
from baryonic semi-leptonic $\beta$-decays under the assumption of 
SU(2) and SU(3) flavor symmetries~\cite{ratcl}. These relate combinations 
of the first moments of the PDFs
to the constants $F$ and $D$ parameterizing the $\beta$-decay rates. We
make use of these constraints in our present analysis; however, 
rather than imposing the exact SU(2) and SU(3) flavor symmetry relations, 
we allow for deviations in our fit within the uncertainty ranges of
the $F,D$ values. Specifically, we set
\begin{eqnarray}
\label{eq:su2}
\Delta \Sigma_u - \Delta \Sigma_d &=&
(F+D)\, [1+\varepsilon_{{\mathrm{SU}}(2)}], \\
\label{eq:su3}
\Delta \Sigma_u + \Delta \Sigma_d-2 \Delta \Sigma_s &=&
(3F-D)\, [1+\varepsilon_{{\mathrm{SU}}(3)}],
\end{eqnarray}
where
\begin{equation}
\label{eq:firstmom1}
\Delta \Sigma_f\equiv\left[ \Delta f^1_i+\Delta \bar{f}^1_i\right] (\mu_0^2) 
\equiv\int_0^1 \left[ \Delta f_i + \Delta \bar{f}_i \right] (x,\mu_0^2)\,dx
\,,
\end{equation}
$\varepsilon_{{\mathrm{SU}}(2,3)}$ parameterize the departures from
exact SU(2) and SU(3) symmetries, and where we use the latest values  
$F+D=1.269\pm0.003$ and $3F-D=0.586\pm0.031$
(see, e.g., Ref.~\cite{ref:hermes-a1pd}). As a practical matter,
we trade the input parameters $N_{u+\bar{u}}$ and $N_{d+\bar{d}}$ in 
Eq.~(\ref{eq:pdf-input}) for $\varepsilon_{{\mathrm{SU}}(2,3)}$ and fit 
the latter. Here the relative uncertainties of $F+D$ and $3F-D$ are assumed 
to represent the typical ranges of $\varepsilon_{{\mathrm{SU}}(2,3)}$; 
we use them to include the deviations from $\varepsilon_{{\mathrm{SU}}
(2,3)}=0$ as additional contributions to $\chi^2$, similarly to the 
case of normalization uncertainties shown in the first term of 
Eq.~(\ref{eq:chi2defnorm}). We note that the relative uncertainties 
of $F+D$ and $3F-D$ are rather modest and may not fully reflect 
the actual breaking of the SU(2) and, in particular, SU(3) symmetries, 
for which larger breaking effects have been discussed in the 
literature~\cite{su3}. This issue may need to be revisited in the 
future. For now we note that as a result of this the PDFs in our fits 
will naturally have a tendency to have relatively small 
$\varepsilon_{{\mathrm{SU}}(2,3)}$. 

Rather than determining also the strong coupling $\alpha_s$ in
the global fit along with the PDFs, we take
$\Lambda_{\mathrm{QCD}}=334.2\,\mathrm{MeV}$ for $n_f=4$ flavors from
the analysis of unpolarized PDFs in Ref.~\cite{ref:mrst}.
The scale dependence of $\alpha_s$ is computed by numerically solving
its renormalization group equation at NLO accuracy.
The charm and bottom quark thresholds are crossed at
$m_c=1.43\,\mathrm{GeV}$ and $m_b=4.3\,\mathrm{GeV}$, respectively.
As already mentioned, the scale evolution equations for the PDFs are
solved analytically in Mellin moment space by explicitly
truncating the solutions at NLO.
Likewise, all observables used in our fit are computed consistently
at NLO accuracy in the $\overline{\mathrm{MS}}$ factorization scheme.
All quarks are treated as massless, and charm and bottom PDFs are turned
on in the evolution at $Q=m_{c,b}$. We note that for all presently available
spin-dependent observables heavy quarks play a negligible role, and,
for the time being, we can safely refrain from introducing more elaborate
variable flavor number schemes which take into
account quark mass effects near threshold (see, e.g., Refs.~\cite{ref:cteq-latest,ref:mstw}).

%%%%%%%%%%%%%%
% TABLE 2
%%%%%%%%%%%%%%
%
\begin{table}[th!]
\caption{\label{tab:para} Parameters $\{a_i^0\}$
describing our optimum NLO ($\overline{\mathrm{MS}}$)
$x\Delta f_i$ in Eq.~(\ref{eq:pdf-input})
at the input scale $\mu_0=1\,\mathrm{GeV}$.}
\begin{ruledtabular}
\begin{tabular}{cccccc}
flavor $i$ &$N_i$ & $\alpha_i$ & $\beta_i$ &$\gamma_i$ &$\eta_i$\\
\hline
$u+\bar{u}$ & 0.677  & 0.692 & 3.34 & -2.18 & 15.87      \\
$d+\bar{d}$ & -0.015 & 0.164 & 3.89 & 22.40 & 98.94      \\
$\bar{u}  $ & 0.295  & 0.692 & 10.0 & 0     & -8.42      \\
$\bar{d}  $ & -0.012 & 0.164 & 10.0 & 0     & 98.94      \\
$\bar{s}  $ & -0.025 & 0.164 & 10.0 & 0     & -29.52     \\
$g             $ & -131.7 & 2.412 & 10.0 & 0     & -4.07      \\
\end{tabular}
\end{ruledtabular}
\end{table}
The parameters $\{a_i^0\}$ representing our best global
fit of polarized parton densities $\Delta f_i$ in Eq.~(\ref{eq:pdf-input}),
henceforth denoted as ``set $S^0$'', are given in Tab.~\ref{tab:para}.
A few additional remarks are in order here.
The currently available data do not fully constrain
the entire $x$ dependence of $\Delta f_i$ imposed in Eq.~(\ref{eq:pdf-input}),
and we are forced to make some restrictions on the parameter
space $\{a_i\}$. For the sea quark and gluon densities we set
$\gamma_i=0$ in Eq.~(\ref{eq:pdf-input}).
This only marginally limits the freedom in the functional form and
still allows nodes. In addition, we tie the
small $x$ behavior, represented by the $\alpha_i$ in Eq.~(\ref{eq:pdf-input}),
of $\Delta u+\Delta \bar{u}$ and $\Delta d+\Delta \bar{d}$ to that of
$\Delta \bar{u}$ and $\Delta \bar{d}$, respectively, which is
reasonable as sea quarks likely dominate in this region.
The parameter $\alpha_{\bar{s}}$ 
always came out close to $\alpha_{\bar{d}}$, so we set them equal.
Since the SIDIS data are not yet sufficient to distinguish
$\Delta s$ from $\Delta \bar{s}$, we assume
$\Delta s=\Delta \bar{s}$ throughout. 
In general, the $x\to 1 $ behavior of all PDFs is rather unconstrained
as is the case even for current sets of unpolarized PDFs.
This is because there are no data sensitive to $x\gtrsim 0.6$.
To avoid problems with the fundamental positivity constraint
\begin{equation}
\label{eq:pos-xsec}
|d\Delta \sigma| \leq d\sigma \ ,
\end{equation}
we make sure that all $\Delta f_i$ vanish at large $x$ at least as fast 
as the $f_i$ of our reference set of unpolarized PDFs~\cite{ref:mrst},
which puts constraints on the $\beta_i$. Choosing any other recent set 
of unpolarized PDFs, like CTEQ6~\cite{cteq6}, does not alter our results.
We note that implementing the positivity constraint at the level
of PDFs \cite{ref:positivity}, i.e.,
\begin{equation}
\label{eq:pos-pdfs}
|\Delta f_i(x,Q^2)| \leq f_i(x,Q^2)
\end{equation}
is strictly valid in LO only, but in the $\overline{\mathrm{MS}}$
scheme is sufficient to guarantee (\ref{eq:pos-xsec}) also in NLO.
The parameters 
$\beta_{u+\bar{u}}$ and $\beta_{d+\bar{d}}$ come out very close
to their corresponding values for the unpolarized case, implying
$(\Delta u+\Delta\bar{u})/(u+\bar{u})\to const.$ as $x\to 1$, and likewise
for $(\Delta d+\Delta\bar{d})/(d+\bar{d})$. 
Since the other $\beta_i$ in Eq.~(\ref{eq:pdf-input}) 
are only very weakly determined 
by the fit, we fix them to their preferred values within the positivity 
constraints whenever we examine uncertainties of the PDFs, in order to 
avoid extremely flat directions in parameter space where $\chi^2$ varies 
only very slowly. Therefore, and as is in general the case in PDF studies, 
our uncertainty estimates for helicity-dependent PDFs are valid only
in the $x$-region explored by experiment. Notice that the near equality 
of $\eta_{d+\bar{d}}$ and $\eta_{\bar{d}}$ is not imposed but a result of the 
fit (in fact, the actual values for these parameters before rounding are
98.9384 and 98.9354, respectively). 

%%%%%%%%%%%%%%%%%
% FIGURE 1
%%%%%%%%%%%%%%%%%
\begin{figure*}[!ht]
\begin{center}
\vspace*{-1cm}
\epsfig{figure=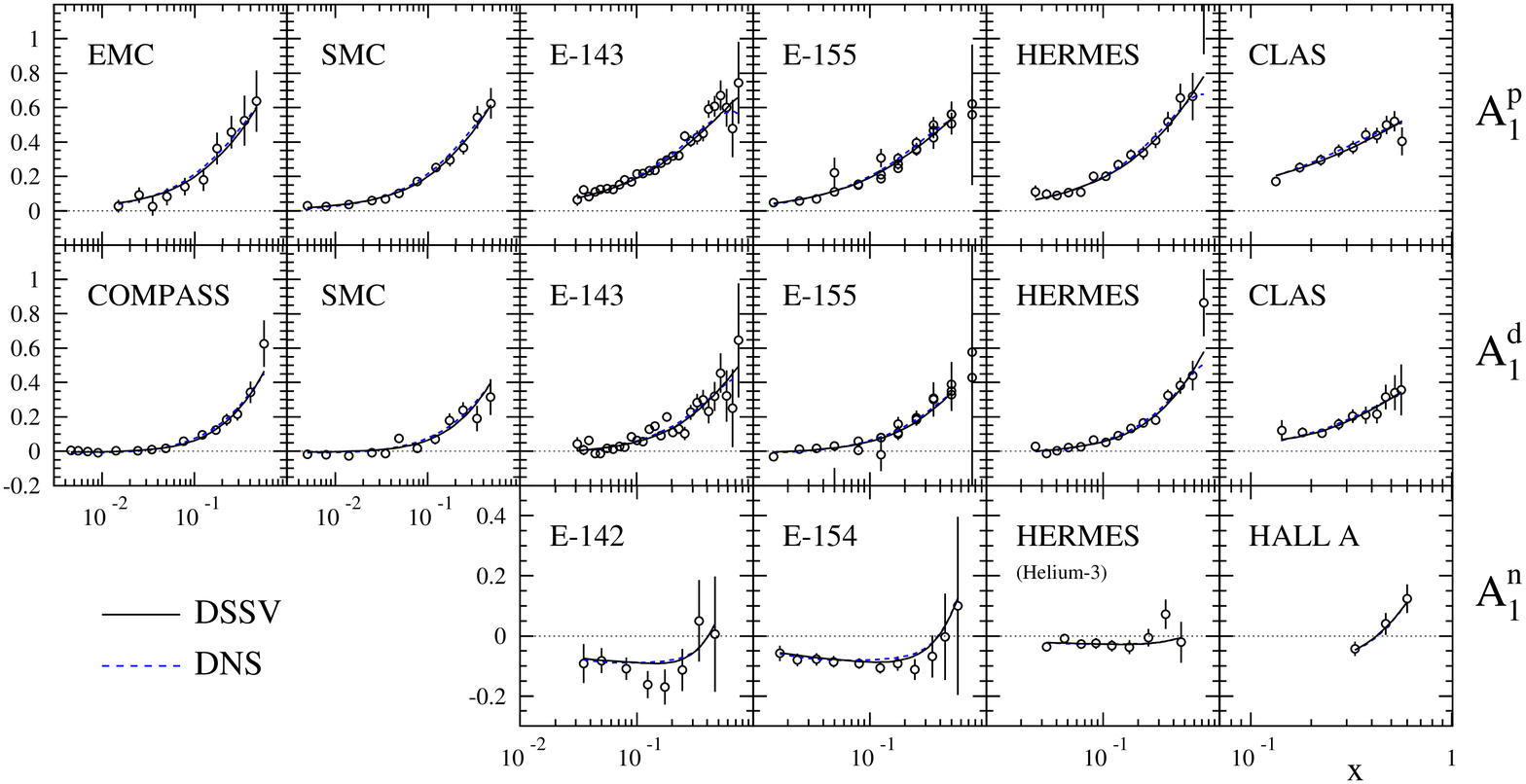,width=0.95\textwidth}
\end{center}
\vspace*{-0.9cm}
\caption{Inclusive DIS spin 
asymmetries~\cite{ref:emc-a1p,ref:smc-a1pd,ref:compass-a1d,ref:e142-a1n,ref:e143-g1pd,
ref:e154-a1n,ref:e155-g1p,ref:e155-g1d,ref:hermes-a1he3-sidisn,ref:hermes-a1pd,
ref:halla-g1n,ref:clas-g1pd} 
compared to the best fit results of our global analysis
(``DSSV'', solid lines), and for the set of polarized PDFs of~\cite{dns} 
(``DNS'', dashed lines). 
\label{fig:dis}}
\vspace*{-0.5cm}
\end{figure*}

In total this leaves us with 19 free parameters in the fit
[or 17, if we fix also $\beta_{u+\bar{u}}$ and $\beta_{d+\bar{d}}$], 
which we include later on also in our uncertainty estimates.
We tried to relax the imposed constraints discussed above, 
but found that present data, i.e., the effective $\chi^2$ function,
are not really sensitive to them.
In Tab.~\ref{tab:para} we have converted the fitted values
for $\varepsilon_{{\mathrm{SU}}(2,3)}$,
defined in Eqs.~(\ref{eq:su2}) and (\ref{eq:su3}),
back to $N_{u+\bar{u}}$ and $N_{d+\bar{d}}$ for convenience.
For the optimal DSSV fit (set $S^0$) we find
\begin{equation}
\label{eq:su2su3-fit}
\varepsilon_{{\mathrm{SU}}(2)}= 0.0011 \,\,,
\,\, \varepsilon_{{\mathrm{SU}}(3)} = -0.0035\,,
\end{equation}
which corresponds to only very minor violations of
the canonical constraints on the first moments
$\Delta \Sigma_u - \Delta \Sigma_d$ and
$\Delta \Sigma_u + \Delta \Sigma_d-2 \Delta \Sigma_s$
assumed in most fits so far. As we have discussed above,
the smallness of $\varepsilon_{{\mathrm{SU}}(2,3)}$ is not really
a surprise in view of the relatively small nominal uncertainty
of the $F+D$ and $3F-D$ values in Eqs.~(\ref{eq:su2}) and (\ref{eq:su3}).
If correct, the small value for $\varepsilon_{{\mathrm{SU}}(3)}$ has 
interesting implications on the behavior of the $\Delta s(x)=\Delta 
\bar{s}(x)$ distribution in the best fit, as we shall see later.

%%%%%%%%%%%%%%%%%%%%%%%%%%%%%%%%%%%%%%%%%%%%%%%%%%%
\subsection{Comparison to fitted data}
\begin{figure*}[!ht]
\begin{center}
\vspace*{-0.3cm}
\epsfig{figure=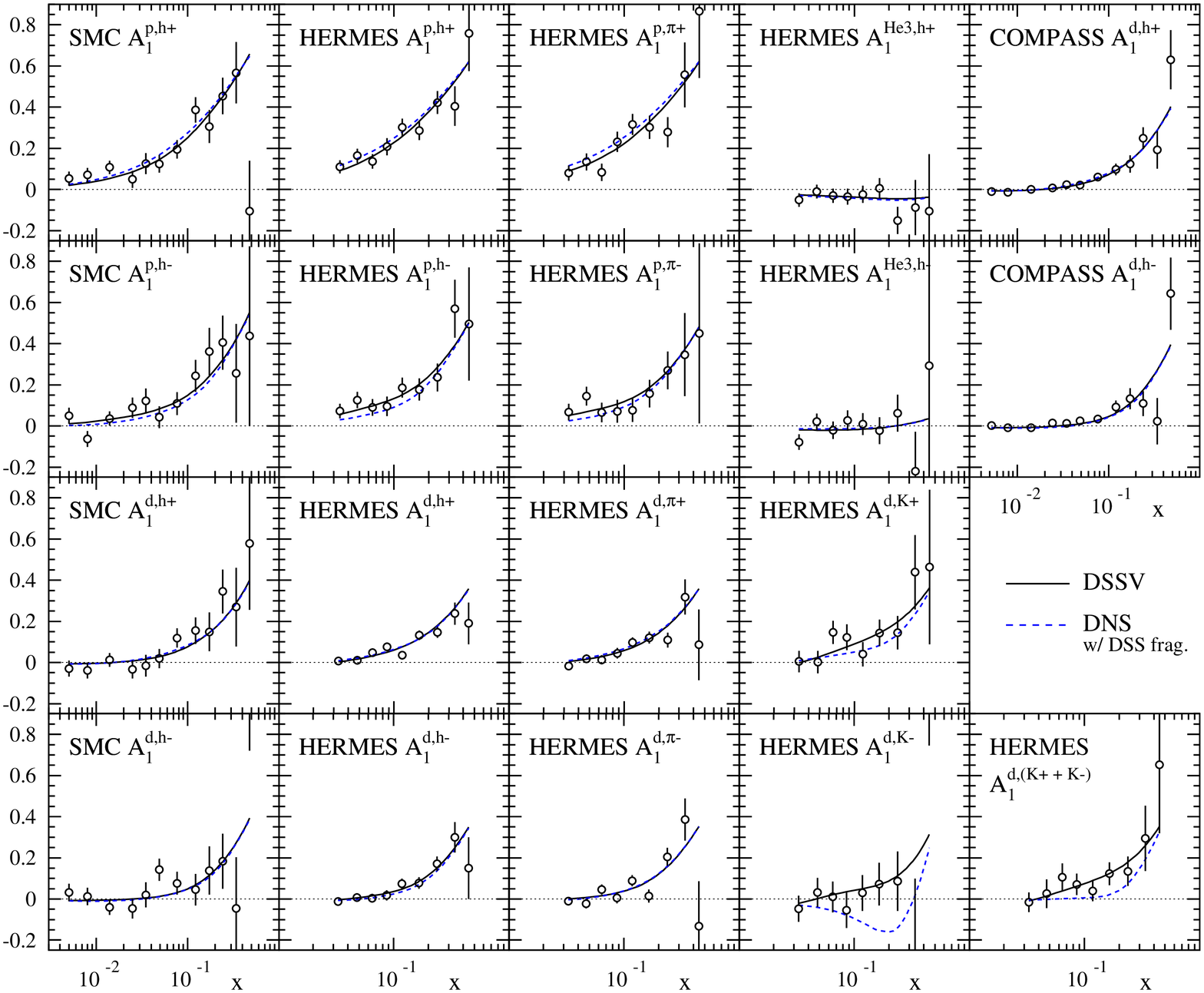,width=0.95\textwidth}
\end{center}
\vspace*{-0.7cm}
\caption{Same as Fig.~\ref{fig:dis}, but for the semi-inclusive DIS 
asymmetries~\cite{ref:smc-sidis,ref:hermes-sidispd,ref:hermes-a1he3-sidisn,ref:compass-sidisd}. 
In all calculations the fragmentation functions of~\cite{ref:dss} have been used.
\label{fig:sidis}}
\vspace*{-0.5cm}
\end{figure*}

%%%%%%%%%%%%%%%%%%%%%%%%%%%%%%%%%%%%%%%%%%%%%%%%%%%
%
The total $\chi^2$ of the best fit $S^0$ is 392.6 for 467 data points
used in our NLO global analysis. We list in Tab.~\ref{tab:chi2table} also
the individual $\chi^2$ values for each experiment. As one can see,
there are only very few cases where the $\chi^2/N_{\mathrm{data}}^{(n)}$
is significantly larger than one. In each case, 
this is due to large fluctuations of some of the data points
in that particular set which are impossible to accommodate in the fit.
Figure~\ref{fig:dis} shows the comparison of our fit to the fitted DIS data,
while the comparison to the SIDIS data is shown in Fig.~\ref{fig:sidis}. 
Notice that in Fig.~\ref{fig:dis} the plots are generically labeled as 
asymmetries ``$A_1$''; however, in the case of the E143, E155, CLAS and 
Hall A data, they actually show the reported structure function ratios and 
are compared to the DSSV estimates for the asymmetries, divided by the  
factor $(1+\gamma^2)$. The overall agreement of the experimental sets in 
the global analysis is excellent. 
All data can be very satisfactorily described by a {\em universal} set of 
polarized PDFs as is assumed by the fundamental factorization theorem.
The agreement with the RHIC $pp$ data is equally good; we have shown
the corresponding comparison in our previous paper~\cite{ref:dssv}
and will come back to it in the next Subsection. 

Figures~\ref{fig:dis} and~\ref{fig:sidis} also show the results obtained
for the set of polarized PDFs of~\cite{dns} in the following 
labelled as ``DNS''. Apart from the fact that
not all of the present data sets were available at the time of~\cite{dns},
a main difference between the two analyses resides in the fragmentation
functions used when including the SIDIS spin asymmetries in the fit. To 
illustrate this point, we have used here the new fragmentation functions 
of~\cite{ref:dss} also for the calculations with the DNS set~\cite{dns}.
As can be seen from Fig.~\ref{fig:sidis}, this leads to significant differences, 
in particular, in the kaon and, to a lesser extent, the isospin sensitive 
(proton target) SIDIS asymmetries. This is mainly due to the strange
fragmentation functions, which directly affect the strange quark polarization,
and also to differences in the light sea quark distributions. Figure~\ref{fig:dis}
shows that there is, however, little difference between the two sets as
far as the inclusive DIS asymmetries are concerned. The changes in the 
strange quark and other sea polarizations are thus compensated here, either mutually 
or by the other parton distributions.

%%%%%%%%%%%%%%%%%%%%%%%%%%%%%%%%%%%%%%%%%%%%%%%%%%%
\subsection{Extracted PDFs, their uncertainties, and their 
physics\label{sec:extracted}}
%%%%%%%%%%%%%%%%%%%%%%%%%%%%%%%%%%%%%%%%%%%%%%%%%%%

%%%%%%%%%%%%%%%%%
% FIGURE 2
%%%%%%%%%%%%%%%%%

%%%%%%%%%%%%%%%%%
% FIGURE 3
%%%%%%%%%%%%%%%%%
\begin{figure}[!ht]
\begin{center}
%\vspace*{-0.6cm}
\epsfig{figure=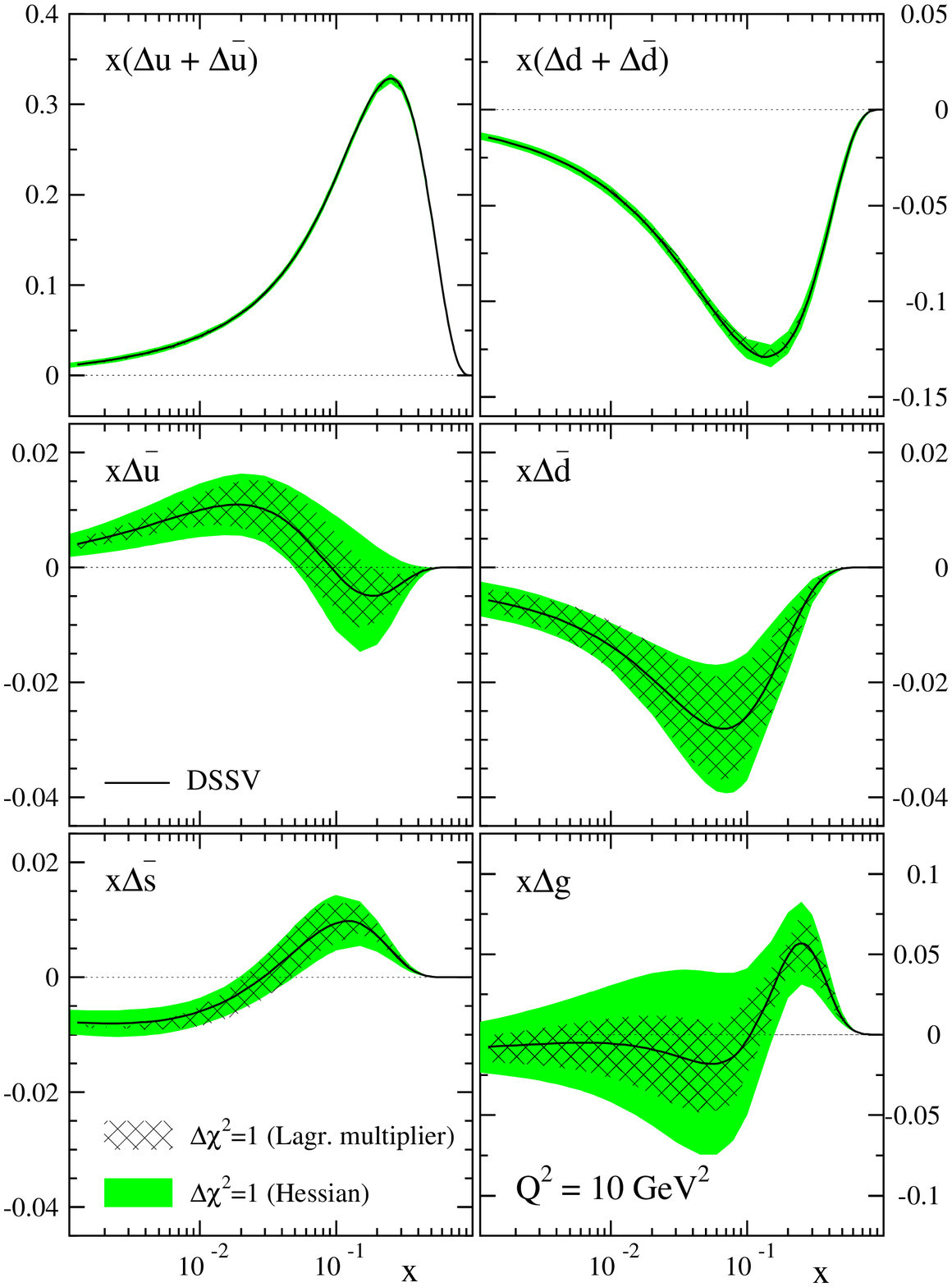,width=0.45\textwidth}
\end{center}
\vspace*{-0.4cm}
\caption{Our polarized PDFs of the proton at $Q^2=10$~GeV$^2$ in the 
$\overline{\mathrm{MS}}$ scheme, along with their $\Delta \chi^2=1$ 
uncertainty bands computed with Lagrange multipliers and the 
improved Hessian approach, as described in the text.
\label{fig:bands}}
%\vspace*{-0.5cm}
\end{figure}

Figure~\ref{fig:bands} shows the extracted polarized PDFs at $Q^2=10$~GeV$^2$, 
along with estimates of their
uncertainties for the Hessian and Lagrange multiplier methods, both for a tolerance of 
$\Delta \chi^2=1$. The results for the Lagrange multiplier method
were already shown in our previous paper~\cite{ref:dssv}
along with a more conservative estimate of the PDF uncertainties based on
$\Delta \chi^2/\chi^2=2\%$.
For this method, the estimates were obtained by 
varying the first moments of the distributions, truncated to the 
region of momentum fractions $0.001\leq x\leq 1$ covered by the data included in the fit.
The distributions corresponding to the maximum variations of the truncated moments
for a given increase $\Delta \chi^2$ were then taken as faithful estimates of the range of 
variation of the PDFs. In the case of the polarized gluon distribution, this procedure was found 
to be not adequate \cite{ref:dssv} because of the fact that there is a significant amount of rather
precise proton-proton collision data constraining the gluon density in however
a relatively narrow region of momentum fraction, $0.05\lesssim x\lesssim 0.2$.
In this way the variations of the gluon distribution's integral in the full region $0.001\leq 
x\leq 1$ tend to produce distributions that favor the variations outside the $pp$ kinematic  
region, misrepresenting the uncertainties inside. In order to circumvent this problem, we performed 
variations of the integral of the gluon distribution in three different $x$ regions,
$0.001\leq x\leq 0.05$, $0.05\leq x\leq 0.2$, and $0.2\leq x\leq 1$, allowing them to 
jointly contribute a change in $\chi^2$ of $\Delta \chi^2=1$. Clearly, the choice for these 
regions and the way they share the increase in $\Delta \chi^2$ is not unique. 
In order to specifically focus on the $x$-region accessed at RHIC, we also performed
a dedicated study of the truncated moment of $\Delta g$ in this region, allowing
variations of $\Delta \chi^2=1$ from this region alone. The results for the truncated
moments of our polarized PDFs,
\begin{equation}
\label{eq:trunc}
\Delta f_i^{1,[x_{\mathrm{min}}\to x_{\mathrm{max}}]}
(Q^2) \equiv \int_{x_{\mathrm{min}}}^{x_{\mathrm{max}}}\Delta f_i(x,Q^2)\,dx \, ,
\end{equation}
are given in Tab.~\ref{tab:hessian}. 
%
%%%%%%%%%%%%%%
% TABLE III
%%%%%%%%%%%%%%
%
\begin{table}[!ht]
\caption{\label{tab:hessian} Truncated first moments $\Delta f_j^{1,[0.001\to 1]}$ at
$Q^2=10\,\mathrm{GeV^2}$ and their uncertainties for $\Delta \chi^2=1$
obtained with the Lagrange multiplier and the Hessian methods. For
future reference, we also recall the results for the Lagrange multiplier
method obtained in~\cite{ref:dssv} under the assumption $\Delta\chi^2/
\chi^2=2\%$, which are to be considered more realistic estimates of
the uncertainties. In the last line, $\Delta g^{\text{RHIC}}$ represents 
the first moment but truncated to $[0.05\to0.2]$.}
\begin{ruledtabular}
\begin{tabular}{cccc}
& \multicolumn{1}{c}{\hspace*{-3mm}Lagr. $\Delta\chi^2=1$} & 
\multicolumn{1}{c}{\hspace*{-3mm}Hessian}& \multicolumn{1}{c}{\hspace*{-3mm}
Lagr. $\Delta\chi^2/\chi^2=2\%$}  \\ \hline \\[-5.5mm] & & & \\
$\Delta u + \Delta\bar{u}$ &  $0.793^{+0.011}_{-0.012}$ &  $0.793{\pm 0.012}$&
$0.793^{+0.028}_{-0.034}$ 
\\[1.5mm]
$\Delta d + \Delta\bar{d}$ & $-0.416^{+0.011}_{-0.009}$ & $-0.416{\pm 0.011}$& 
$-0.416^{+0.035}_{-0.025}$ \\[1.5mm]
$\Delta\bar{u}$            &  $0.028^{+0.021}_{-0.020}$ &  $0.028{\pm 0.022}$& 
$0.028^{+0.059}_{-0.059}$ \\[1.5mm]
$\Delta\bar{d}$            & $-0.089^{+0.029}_{-0.029}$ & $-0.089{\pm 0.029}$& 
$-0.089^{+0.090}_{-0.080}$ \\[1.5mm]
$\Delta\bar{s}$            & $-0.006^{+0.010}_{-0.012}$ & $-0.006{\pm 0.012}$&
$-0.006^{+0.028}_{-0.031}$  \\[1.5mm]
$\Delta\Sigma$             &  $0.366^{+0.015}_{-0.018}$ &  $0.366{\pm 0.017}$& 
$0.366^{+0.042}_{-0.062}$ \\[1.5mm]
$\Delta g$                 &  $0.013^{+0.106}_{-0.120}$ &  $0.013{\pm 0.182}$&
$0.013^{+0.702}_{-0.314}$ \\[1.5mm]
$\Delta g^{\mathrm{RHIC}}$  &  $0.005^{+0.051}_{-0.058}$ & $0.005{\pm 0.056}$&
$0.005^{+0.129}_{-0.164}$ \\[0.5mm]
\end{tabular}
\end{ruledtabular}
\end{table}

\begin{figure}[!ht]
\begin{center}
%\vspace*{-0.6cm}
\epsfig{figure=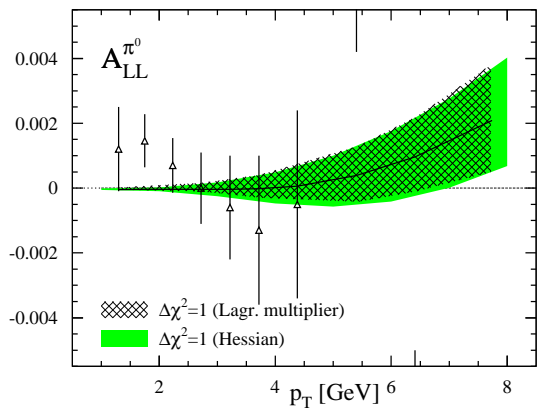,width=0.45\textwidth}
\end{center}
\vspace*{-0.7cm}
\caption{Uncertainties of the calculated $A_{LL}^{\pi^0}$ at RHIC in our global fit, 
computed
using both the Lagrange multiplier and the Hessian matrix techniques. We also show 
the corresponding PHENIX data~\cite{ref:phenix-pion-run6}.
\label{fig:rhic-all-hessian}}
\vspace*{-0.5cm}
\end{figure}

Inspection of Fig.~\ref{fig:bands} and Tab.~\ref{tab:hessian} reveals that the Hessian and
the Lagrange multiplier methods yield fairly similar $\Delta \chi^2=1$  uncertainties, except
for the spin-dependent gluon distribution, 
for which the Lagrange multiplier approach yields a still significant but generally smaller uncertainty 
than the one predicted by the Hessian method using Eq.~(\ref{eq:obserror-hessian}).
The agreement between the two often becomes better 
when the observable is better constrained by the data, as is the case for the integral of 
$\Delta g$ over only the $x$-range probed at RHIC, or for the actual physical observables
that determine $\Delta g$. As an example, in Fig.~\ref{fig:rhic-all-hessian} we show the
estimated uncertainties for the double-longitudinal spin asymmetry,
\begin{equation}
A_{LL}\equiv\frac{\sigma^{++}-\sigma^{+-}}{\sigma^{++}+\sigma^{+-}}
\label{eq:alldef}
\end{equation}
for $pp\to \pi^0X$ at RHIC, where the superscripts denote the helicities of
the incoming protons, computed 
with both the Lagrange multiplier and the improved Hessian approaches. As can be seen,
the two give very similar results. This feature can be traced back to correlations between the
parameters, in the sense that some of them can compensate variations forced in the others. 
We note that such kinds of correlations are fully accounted for in the Lagrange 
multiplier approach, whereas it is not generally clear how well are they represented by the 
approximated Hessian matrix. We shall investigate the distinctive features between the two 
methods later, but will focus first on the physics aspects related to our extracted 
polarized PDFs. 
%
%%%%%%%%%%%%%%%%%
% FIGURE 4
%%%%%%%%%%%%%%%%%

%
%%%%%%%%%%%%%%%%%%%%
% TABLE IV
%%%%%%%%%%%%%%%%%%%%
\begin{table*}[!htb]
\caption{\label{tab:mom} Truncated 
first moments, $\Delta f_i^{1,[0.001\to 1]}$, and full ones, $\Delta f_i^1$, 
of our polarized PDFs at various $Q^2$.}
\begin{ruledtabular}
\begin{tabular}{ccccccccc}
$x$-range in Eq.~(\ref{eq:trunc}) & $Q^2$ [GeV$^2$] &$\Delta u + \Delta\bar{u}$ 
&$\Delta d + \Delta\bar{d}$ &
$\Delta\bar{u}$  & $\Delta\bar{d}$ & $\Delta\bar{s}$ &  $\Delta g$  &
$\Delta\Sigma$   \\ \hline
0.001-1.0 &  1 &    0.809 & -0.417  & 0.034 & -0.089&  -0.006 & -0.118
& 0.381 \\
  &  4 &    0.798 & -0.417  & 0.030 & -0.090&  -0.006 & -0.035 & 0.369\\
  & 10&    0.793 & -0.416  & 0.028 & -0.089&  -0.006 &  0.013 & 0.366 \\
  & 100&  0.785 & -0.412  & 0.026 & -0.088&  -0.005 &  0.117 & 0.363 \\ \hline
0.0-1.0 & 1 &      0.817  &-0.453 &  0.037 & -0.112 & -0.055 & -0.118
&  0.255\\
   &4 &      0.814  &-0.456 &  0.036 & -0.114 & -0.056 & -0.096 &  0.245\\
   &10&     0.813  &-0.458 &  0.036 & -0.115 & -0.057 & -0.084 &  0.242\\
   &100&   0.812  &-0.459 &  0.036 & -0.116 & -0.058 & -0.058 &  0.238\\ 
\end{tabular}
\end{ruledtabular}
\end{table*}
Table~\ref{tab:mom} shows the evolution of the central values for the truncated 
first moments $\Delta f_i^{1,[0.001\to 1]}$ with $Q^2$. 
$\Delta \Sigma$ denotes the quark singlet combination, i.e., the sum
of all quarks and anti-quarks. We also show the
evolution of the full first moments $\Delta f_i^1$. These obviously rely on an 
extrapolation of the PDFs to $x$-values outside the measured region, 
and it is difficult to estimate the uncertainty associated with this. 

\noindent{\em Total up and down distributions}: $\Delta u+\Delta \bar{u}$ and $\Delta d+
\Delta \bar{d}$, which inclusive DIS probes primarily, are the by far 
best determined distributions. Their uncertainty bands are very narrow, 
see Fig.~\ref{fig:bands}, and also our
results agree very well with the determinations in previous analyses~\cite{grsv,bb,lss,aac,dns}.
We note that recent lattice QCD results~\cite{latt} of the full 
first moments $\Delta \Sigma_u\equiv \Delta u^1+\Delta \bar{u}^1$ and 
$\Delta \Sigma_d\equiv \Delta d^1+\Delta \bar{d}^1$ (albeit excluding
disconnected diagrams) also agree very well with the values we extract, 
which may shed light on the validity of assumed extrapolations of the 
parton distribution functions to small $x$. 

%%%%%%%%%%%%%%%%%
% FIGURE 5
%%%%%%%%%%%%%%%%%
\begin{figure}[!ht]
\begin{center}
%\vspace*{5cm}
\epsfig{figure=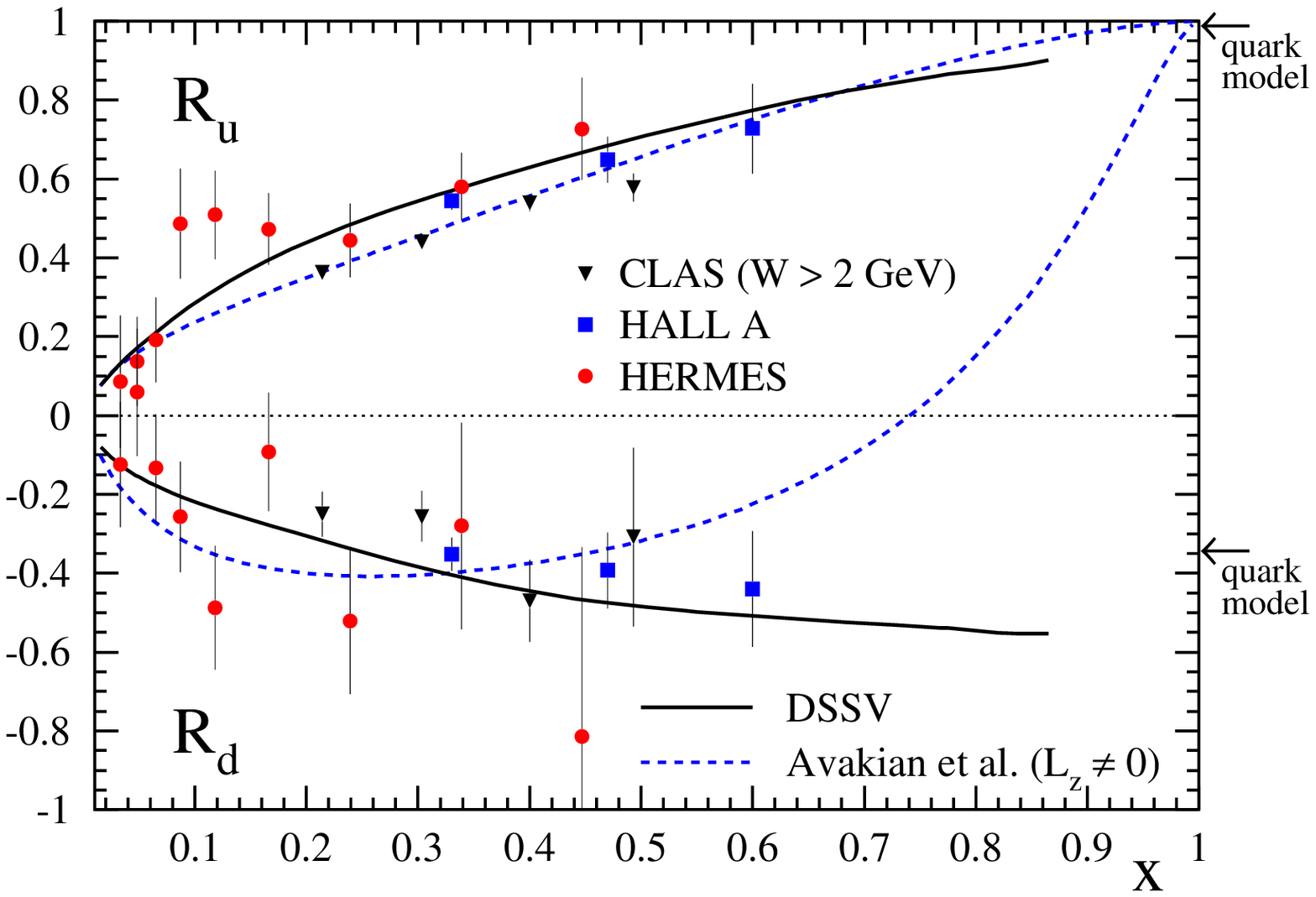,width=0.5\textwidth}
\end{center}
\vspace*{-0.7cm}
\caption{High-$x$ behavior of our $R_u=(\Delta u+\Delta \bar{u})/
(u+\bar{u})$ and $R_d=(\Delta d+\Delta \bar{d})/
(d+\bar{d})$, shown by the solid curves, along with the data
from~\cite{ref:halla-g1n,ref:clas-g1pd,ref:hermes-sidispd}, 
as shown in~\cite{deur}. The dashed lines present the predictions 
based on power counting and perturbative QCD, taking into account 
nonzero orbital angular momentum Fock states~\cite{deur}. We note that
these results, as well as some of the experimental data, are
for the ratios $\Delta u/u$ and $\Delta d/d$ rather than for $R_u$
and $R_d$, which, however, makes little difference for the 
large $x$ values we consider here. The arrows
show the expectations for $\Delta u/u$ and $\Delta d/d$ 
in relativistic constituent quark models~\cite{isgur}. 
\label{fig:largex}}
%\vspace*{0.5cm}
\end{figure}
We have mentioned earlier that in our fit $R_u\equiv 
(\Delta u+\Delta \bar{u})/(u+\bar{u})$ and 
$R_d\equiv (\Delta d+\Delta \bar{d})/(d+\bar{d})$ 
become constant in the ``valence region'' as $x\to 1$, where the sea
quark contributions become small. Figure~\ref{fig:largex} 
shows the ratios $R_u,R_d$ along with the most relevant experimental data. 
The information at the highest values of $x$ comes
from the Jefferson Laboratory Hall-A experiment~\cite{ref:halla-g1n}. As 
one can see, our $R_u$ goes to unity at high $x$, which is consistent 
with expectations in relativistic 
constituent quark models~\cite{isgur}, but also in perturbative QCD, using power counting
and hadron helicity conservation~\cite{fj}. We furthermore find that 
$R_d$ remains negative 
in the region where it is constrained by data and presently shows no tendency to turn 
towards $+1$ at high $x$. The latter behavior would be expected for the pQCD
based models. We note that it has recently been 
argued~\cite{deur} that the upturn of $R_d$ in such models 
could set in only at relatively high $x$, due to the presence 
of valence Fock states of the nucleon with nonzero orbital angular momentum that produce 
double-logarithmic contributions $\sim \ln^2(1-x)$ in the limit of $x\to 1$ on top
of the nominal power behavior. The corresponding expectation is also shown in the figure.
In contrast to this, relativistic constituent quark models predict 
$R_d$ to tend to $-1/3$
as $x\to 1$, perfectly consistent with the present data.

%%%%%%%%%%%%%%%%%
% FIGURE 6
%%%%%%%%%%%%%%%%%
\begin{figure}[!ht]
\begin{center}
\vspace*{-0.6cm}
\epsfig{figure=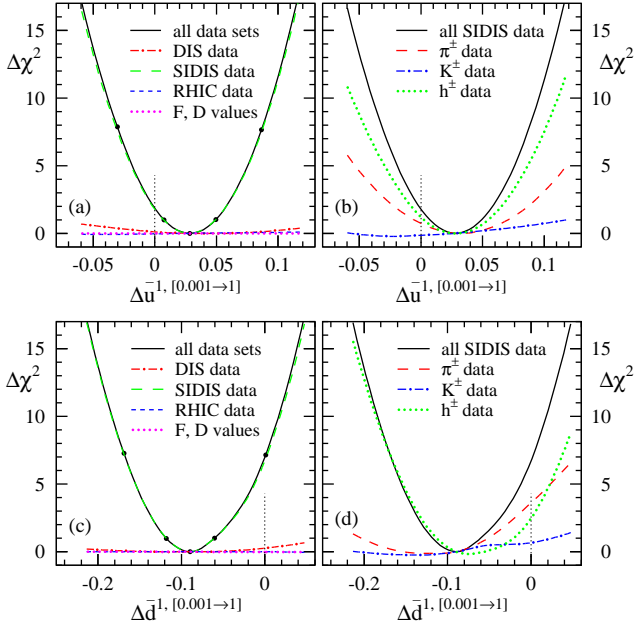,width=0.5\textwidth}
\end{center}
\vspace*{-0.7cm}
\caption{{\bf{(a)}} The $\chi^2$ profiles and partial contributions $\Delta \chi^2$  
of the various types of data sets for variations of the truncated first moment 
$\Delta \bar{u}^{1,[0.001\to 1.0]}$ at $Q^2=10$~GeV$^2$. 
{\bf{(b)}} Partial contributions $\Delta \chi^2$ for the various hadrons \
produced in SIDIS. {\bf{(c)}} and {\bf{(d)}}: same as (a) and (b), 
but for $\Delta \bar{d}^{1,[0.001\to 1.0]}$.
\label{fig:ubar}}
%\vspace*{0.5cm}
\end{figure}

\noindent{\em Light sea quark polarizations:}  
The light sea quark and anti-quark distributions turn out to be better constrained now
than in previous analyses~\cite{dns}, thanks to the advent of more precise 
SIDIS data~\cite{ref:smc-sidis,ref:hermes-sidispd,ref:hermes-a1he3-sidisn,ref:compass-sidisd} 
and of the new set of fragmentation functions~\cite{ref:dss} that describes the 
observables well in the unpolarized case. Figure~\ref{fig:ubar} shows the changes
in $\chi^2$ of the fit as functions of the truncated first moments $\Delta 
\bar{u}^{1,[0.001\to 1]}, \Delta \bar{d}^{1,[0.001\to 1]}$ defined in Eq.~(\ref{eq:trunc}), 
obtained for the Lagrange multiplier method. On the 
left-hand-side, Figs.~\ref{fig:ubar} (a), (c), 
we show the effect on the total $\chi^2$, as well as on the $\chi^2$ values for the individual 
contributions from DIS, SIDIS, and RHIC $pp$ data and from the $F, D$ values. It is 
evident that the SIDIS data completely dominate the changes in $\chi^2$. 
On the r.h.s.\ of the plot, Figs.~\ref{fig:ubar} (b), (d), 
we further split up $\Delta \chi^2$ from SIDIS 
into contributions associated with the spin asymmetries in 
charged pion, kaon, and unidentified hadron production. One can see that the latter dominate,
closely followed by the pions. The  kaons have negligible impact here. 
For $\Delta \bar{u}^{1,[0.001\to 1]}$, charged hadrons and pions are very consistent, as far as 
the location of the minimum in $\chi^2$ is concerned.  For $\Delta \bar{d}^{1,[0.001\to 1]}$
there is some slight tension between them, although it is within the tolerance of the 
fit.

\begin{figure}[!ht]
\begin{center}
\vspace*{-1cm}
\epsfig{figure=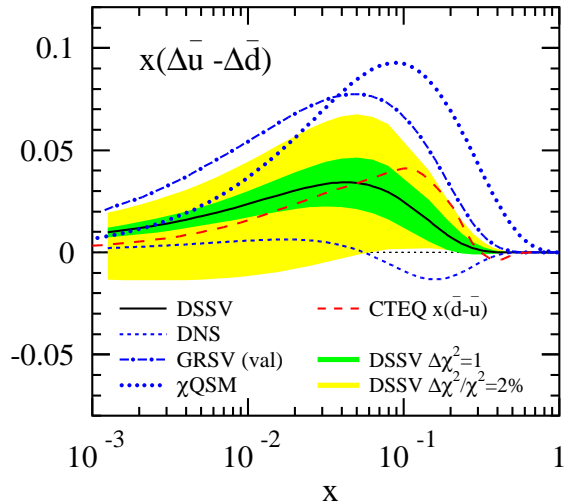,width=0.52\textwidth}
\end{center}
\vspace*{-0.7cm}
\caption{The difference between $x\Delta \bar{u}$ and 
$x\Delta \bar{d}$  at $Q^2=10$~GeV$^2$, along with the uncertainty
bands for $\Delta \chi^2=1$ and $\Delta \chi^2/\chi^2=2\%$. 
The dot-dashed and dotted lines show the predictions of
the valence scenario of~\cite{grsv} and the chiral quark
soliton model of~\cite{models1}, respectively. We also show
the result obtained in an earlier global analysis~\cite{dns} 
of DIS and SIDIS data (light dotted), for which the fragmentation
functions of~\cite{ref:dss} were not yet available. The dashed
line displays for comparison the flavor asymmetry $x(\bar{d}-
\bar{u})$ in the spin-averaged case, using the PDFs of~\cite{cteq6}.
\label{fig:ubardbar}}
\vspace*{0.5cm}
\end{figure}

Of particular physics interest is a possible flavor symmetry breaking in the 
light sea, i.e., $\Delta \bar{u}\neq \Delta \bar{d}$, given the well-established 
significant difference between $\bar{u}$ and $\bar{d}$ in the spin-averaged 
case~\cite{ref:cteq-latest,ref:mstw}. Figure~\ref{fig:bands} indeed clearly points to 
a largely positive $\Delta \bar{u}$ distribution, but a negative (and larger) 
$\Delta \bar{d}$.
Figure~\ref{fig:ubardbar} specifically shows the difference 
$x(\Delta \bar{u}-\Delta \bar{d})$, which is positive within uncertainties. 
Note that we show both the $\Delta \chi^2=1$ and the more 
conservative $\Delta \chi^2/\chi^2=2\%$ uncertainty bands here. 

The pattern of symmetry breaking in the light anti-quark sea polarizations 
shown by Figs.~\ref{fig:bands} and~\ref{fig:ubardbar} has been predicted at
least qualitatively by a number of models of nucleon structure. 
A simple intuitive consideration of the Pauli principle roughly gives
the observed picture: if valence-$u$ quarks primarily spin along 
the proton spin direction, $u\bar{u}$ pairs in the sea will tend to 
have the $u$ quark polarized opposite to the proton. Hence, if such 
pairs are in a spin singlet, one expects $\Delta\bar{u} > 0$ and, 
by the same reasoning, $\Delta\bar{d} < 0$. Expectations based
on the Pauli principle have been made quantitative in~\cite{models}
and the ``valence'' scenario of~\cite{grsv}, and the resulting predictions 
are shown by the dot-dashed line in Fig.~\ref{fig:ubardbar}. They tend
to lie somewhat higher than our extracted $\Delta \bar{u}-\Delta 
\bar{d}$, but are certainly qualitatively consistent, given the still relatively
large uncertainties. The same is true for the case of the chiral
quark soliton model~\cite{models1}, represented by the dotted line in 
the figure. Within the large-$N_c$ limit of QCD on which this
model is based, one in fact expects $|\Delta\bar{u}-\Delta\bar{d}|
> |\bar{u}-\bar{d}|$. As comparison of our extracted $x(\Delta\bar{u}-
\Delta\bar{d})$ with the result of~\cite{cteq6} for $x (\bar{d}-\bar{u})$ 
in Fig.~\ref{fig:ubardbar} shows, one can presently not yet decide
whether this expectation is fulfilled. Predictions for $\Delta \bar{u}-\Delta 
\bar{d}$ have also been obtained within meson cloud models~\cite{models2}; 
it has been found in~\cite{models2a} that also here a flavor 
asymmetry of similar size is possible. 
Finally, also statistical parton models~\cite{soffer,models3}
obtain a similar size of $\Delta \bar{u}-\Delta \bar{d}$. We note that
predictions for the individual $\Delta \bar{u}$ and $\Delta \bar{d}$,
where available, agree on $\Delta \bar{u}>0$, $\Delta \bar{d}<0$,
consistent with our results in Fig.~\ref{fig:bands}, but may differ
in the relative size of the distributions. For example, the results
of~\cite{grsv,models} have $|\Delta \bar{d}|>\Delta \bar{u}$, as in 
Fig.~\ref{fig:bands}, while the statistical models find the two
distributions to be of more equal absolute size.

%%%%%%%%%%%%%%%%%
% FIGURE 7
%%%%%%%%%%%%%%%%%

\noindent{\em Strange quark polarization}: The polarization of strange 
quarks has been a focus since the very beginning of the proton spin crisis.
The reason is that in the parton model and assuming SU(3) symmetry 
(see Sec.~\ref{secIIIa}) one has
\begin{equation}
\label{eq:sigstr}
\Delta \Sigma \equiv \Sigma_u+\Sigma_d+\Sigma_s=
(3F-D) + 3 \Delta \Sigma_s \ ,
\end{equation}
where the $\Delta \Sigma_f$ are as defined in Eq.~(\ref{eq:firstmom1})
but now for arbitrary scale $Q$,
and $\Delta \Sigma$ is the total quark and anti-quark spin contribution
to the proton spin. If the latter is found to be small experimentally,
$\Delta\Sigma\sim 0.25$, the implication is that strange quarks make a 
significant negative contribution to the proton spin. Indeed, most fits 
to only inclusive DIS data have preferred a large and negative strange
quark polarization. The same was found in Ref.~\cite{dns}, even though 
here the SU(3) flavor symmetry was not enforced. 

At variance with these results, the best fit in our present analysis 
has a polarized strange distribution $\Delta s$ that is positive
at large $x$, but negative at small momentum fractions. Before
we discuss the origin and significance of this result, we note 
that a prerequisite for it is that we have adopted 
a more flexible parameterization for the strange quark distribution in
this work, which permits a node. This is again in contrast with the 
previous fits in which the initial $\Delta s$ always had the same 
sign for all $x$. We have assumed however $\Delta s = \Delta \bar{s}$, 
since the fit is unable to discriminate strange quarks from anti-quarks.
This is really an assumption: unlike the spin-averaged
case where the distributions $s$ and $\bar{s}$ will be rather
similar (the integral of $s-\bar{s}$ has to vanish), there is a priori
no need for $\Delta s$ and $\Delta \bar{s}$ to have the same size
or even the same sign.

%%%%%%%%%%%%%%%%%
% FIGURE 8
%%%%%%%%%%%%%%%%%
\begin{figure}[!ht]
\begin{center}
\vspace*{-0.6cm}
\epsfig{figure=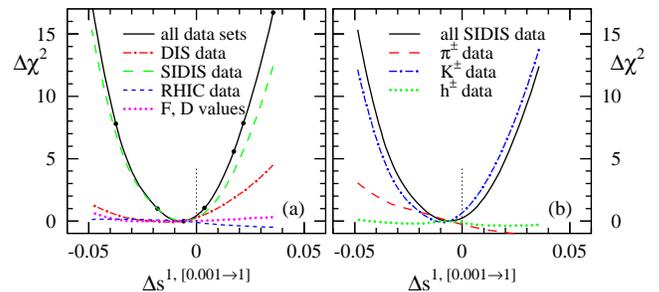,width=0.5\textwidth}
\end{center}
\vspace*{-0.7cm}
\caption{Same as Fig.~\ref{fig:ubar}, but for the truncated first
moment of the polarized strange distribution $\Delta s^{1,[0.001-1.0]}$. 
\label{fig:strange}}
\vspace*{0.5cm}
\end{figure}

Qualitatively, the main features of our extracted strange sea 
distribution arise in the following way: the (kaon) SIDIS data, within
the leading-twist framework we employ, turn out to prefer a small 
and likely positive $\Delta s$ at medium $x$, while inclusive DIS and 
the constraints from $\beta$-decays demand a negative integral of 
$\Delta s$ and so force $\Delta s$ to turn negative at low $x$. 
Given the importance of $\Delta s$, we address these constraints 
and their significance and implications in more detail in the 
following. 

We start by we analyzing the behavior of the truncated first moment,
$\Delta s^{1,[0.001\to 1]}$, around the minimum defining 
the best fit. Figure~\ref{fig:strange} shows the increase of $\chi^2$ 
of the fit against variations of $\Delta s^{1,[0.001\to 1]}$, along 
with the partial contributions of the various data sets. Evidently, 
the best fit has a truncated moment close to zero and only slightly 
negative, as we also saw in Tab.~\ref{tab:hessian}. The shape of 
$\Delta \chi^2$ around the minimum is dominated by the SIDIS data, 
and here primarily by the data for kaon production. All other data sets, 
pion SIDIS, inclusive DIS, and RHIC $pp$ data, play less important roles, 
as expected (here one has of course to keep in mind 
that the impact of {\it individual} data sets seen in the Lagrange 
multiplier scans is always estimated in the ``presence'' of the other 
data sets, and therefore should not be construed as an independent fit 
result). As can be seen from Fig.~\ref{fig:strange} and 
Tab.~\ref{tab:hessian}, the truncated moment of $\Delta s$ remains
close to zero if changes of $\Delta \chi^2=1$ are permitted. For 
the more realistic choice $\Delta \chi^2/\chi^2=2\%$, one finds
that a much larger range of $\Delta s^{1,[0.001\to 1]}$ is allowed,
extending from significantly negative to positive values. The size and even
the sign of the considered truncated moment are, therefore, presently not 
well constrained. Nonetheless, there is a trend for $\Delta s(x)$ to be 
positive at medium $x\sim 0.1$, even for the choice $\Delta \chi^2/
\chi^2=2\%$ (see Fig.~2 of~\cite{ref:dssv}). We note that the COMPASS 
experiment has recently presented a LO extraction of the polarized
strange distribution from their kaon SIDIS data~\cite{compassrw}, 
which are not yet included in this work. These are consistent
with small strangeness polarization down to below $x\sim 0.01$,
but also allow a significantly negative $\Delta s$ at $x\sim 0.005$. 
Furthermore, an extraction of the integral of $\Delta s$ over the range $0.02\le 
x\le 0.6$ by the HERMES collaboration \cite{ref:hermes-strange} 
yields $0.037\pm0.019\pm0.027$, consistent with our result. 

We stress that beyond the ``data-driven'' uncertainties that we find for the 
polarized strangeness distribution, there could well be effects
that are outside the leading-twist framework we are using here 
and that may have a significant impact on the extracted $\Delta s$. 
Given the generally low hadron multiplicities in kaon events in the 
present SIDIS measurements, it is not ruled out that the kaon SIDIS data 
are affected by higher twist contributions and not suited for an
extraction of leading twist strangeness distributions. We note that 
the information on the parton-to-kaon fragmentation functions 
in~\cite{ref:dss} also primarily comes from unpolarized kaon SIDIS data 
and would not be reliable in the latter case either. A recent 
determination of the unpolarized strange distribution in the nucleon 
by HERMES from their SIDIS multiplicities shows an unexpected
shape of the distribution~\cite{ref:hermes-strange}. SIDIS measurements at 
smaller $x$, as well as at presently available $x$, but higher $Q^2$, 
will likely be vital for clarifying these issues. These would become 
available at an electron-ion collider~\cite{raju}.

As can be seen from Fig.~\ref{fig:strange}, the effects due to SU(2) and 
SU(3) flavor symmetry breaking in usage of the baryon semi-leptonic 
decay data, see Eqs.~(\ref{eq:su2}),(\ref{eq:su3}), have only a very limited 
impact on the truncated moment of $\Delta s$. This, however, changes
dramatically when the full first moment of $\Delta s$ is considered,
i.e., the contribution to its integral from $x<0.001$. This region is
presently not constrained by any DIS or SIDIS data, but we remind
the reader that the breaking parameters $\varepsilon_{{\mathrm{SU}}(2,3)}$ 
come out very small, see Eq.~(\ref{eq:su2su3-fit}), in our analysis,
as a result of the relatively small nominal uncertainty in the $F,D$ 
values, as we discussed in Sec.~\ref{secIIIa}. This implies that
the strange sea distribution will have a large and negative total first
moment, $\Delta \Sigma_s=\Delta s^1+\Delta \bar{s}^1=-0.114$ as
seen from Tab.~\ref{tab:mom}, which in turn can only occur
if the distribution shows a sign change to negative values at
small $x$, visible in Fig.~\ref{fig:bands}. 

It will clearly be important in the future to better understand the 
strange contribution to nucleon spin structure. If the full first
moment $\Delta \Sigma_s$ is small, SU(3) symmetry in relating 
hyperon $\beta$-decays to nucleon spin structure would have to 
be broken at the $40\%$ level or so, which is not ruled 
out~\cite{ratcl,su3}. If, on the other hand, SU(3) symmetry is not
broken significantly, the implication is that either $\Delta s$ 
turns large and negative at small $x$, as in our fit, or that the present
kaon SIDIS data do not allow a reliable extraction of $\Delta s(x)$. 
On the theoretical side, there have been very recent lattice determinations 
of the integral $\Delta \Sigma_s$~\cite{latt1}, which point to small values.
Models of nucleon structure, on the other hand, have led to quite 
varied predictions for the integral of $\Delta s$, ranging from
small to large negative values~\cite{models-s}. We note that the ``valence scenario'' 
of~\cite{grsv} has a first moment of the polarized strange distribution very close 
to zero, which is consequently at the expense of significant violation of the SU(3) 
flavor symmetry relation in Eq.~(\ref{eq:su3}). We finally 
stress that the size of $\Delta s$ is 
not a topic of interest just for nucleon spin structure enthusiasts: 
as was pointed out recently~\cite{darkm}, the uncertainty in 
$\Delta \Sigma_s$ provides the single largest uncertainty in predictions 
of the spin-dependent elastic scattering cross sections of supersymmetric 
dark matter particles on protons and neutrons.

%%%%%%%%%%%%%%%%%
% FIGURE 9
%%%%%%%%%%%%%%%%%
\begin{figure}[!ht]
\begin{center}
\vspace*{-0.6cm}
\epsfig{figure=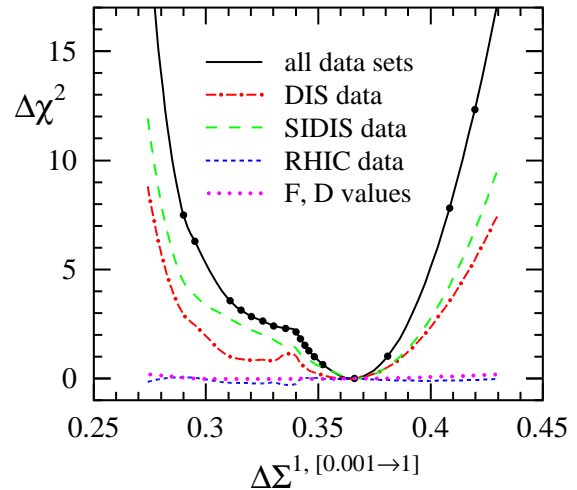,width=0.5\textwidth}
\end{center}
\vspace*{-0.7cm}
\caption{Same as Fig.~\ref{fig:ubar}, but for the truncated first
moment of the quark singlet distribution $\Delta \Sigma^{1,[0.001-1.0]}$. 
\label{fig:sigma}}
\vspace*{0.5cm}
\end{figure}

\noindent{\em Total quark and anti-quark spin contribution $\Delta \Sigma$:}
In Fig.~\ref{fig:sigma} we show the $\chi^2$ profile associated with 
variations of the truncated moment of the quark singlet distribution,
$\Delta\Sigma^{1,[0.001\to 1]}\equiv 
\int_{0.001}^1 dx[\Delta u+\Delta \bar{u}+\Delta d+\Delta \bar{d}+
\Delta s+\Delta \bar{s}]$, at $Q^2=10$~GeV$^2$. As expected, the main 
constraints come from the DIS and SIDIS data. The value for the 
truncated first moment obtained in the best fit is significantly 
higher than that for the full first moment given in Tab.~\ref{tab:mom},
which is a manifestation of the large negative contribution from 
strange quarks and anti-quarks that arises in our fit at small $x$.
Thus, keeping in mind the discussion about strangeness above, 
we conclude that if SU(3) flavor symmetry in relating 
hyperon $\beta$-decays to nucleon spin structure is strongly 
broken, $\Delta \Sigma$ would be as large as $\sim 0.36$ or so,
whereas it will be about $30\%$ smaller if SU(3) holds well and 
the first strange moment $\Delta \Sigma_s$ turns out to be large 
and negative. We note that such lower values of $\Delta\Sigma
\sim 0.24$ or so have usually been obtained in previous 
analyses relying on the use of SU(3) symmetry~\cite{grsv,bb,lss,aac,dns}. 
In any case, $\Delta\Sigma$ is certainly much smaller than the typical
expectation of $\Delta\Sigma\gtrsim 0.6$ in quark models.

%%%%%%%%%%%%%%%%%
% FIGURE 10
%%%%%%%%%%%%%%%%%
\begin{figure}[!ht]
\begin{center}
%\vspace*{5cm}
\epsfig{figure=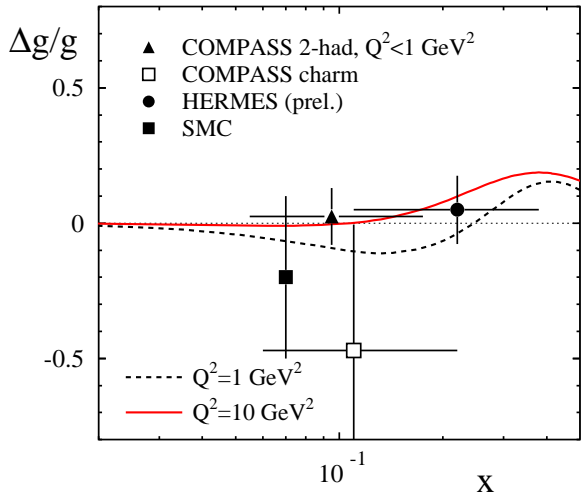,width=0.5\textwidth}
\end{center}
\vspace*{-0.7cm}
\caption{Comparison of $\Delta g/g$ for our best fit, at two
representative $Q^2$, to the extracted $\Delta g/g$ from photon-gluon
fusion processes investigated by SMC~\cite{ref:smc}, 
HERMES~\cite{ref:hermes}, and 
COMPASS~\cite{ref:compass-2had,ref:compass-charm}. 
These data were not included in our global analysis since a consistent
NLO framework is not available at present. 
\label{fig:dgdis}}
%\vspace*{0.5cm}
\end{figure}

\noindent{\em Spin-dependent gluon distribution $\Delta g$:}
We have already noted in our DSSV paper~\cite{ref:dssv} that
the polarized gluon distribution $\Delta g(x,Q^2)$ comes out rather 
small in the presently accessed range of momentum fraction $x$, and 
prefers to have a node. 
At variance with the findings of Ref.~\cite{aac}, we do not find any
non-overlapping best-fit solutions with gluon polarizations of
opposite signs. This duplicity is readily excluded by the RHIC $pp$ data.
The RHIC data in fact turn out to play a crucial 
role in constraining $\Delta g$ \cite{ref:dssv}. The result is 
shown again in Fig.~\ref{fig:bands}. We do not repeat the 
plot of the $\chi^2$ profile as a function of the truncated
first moment of $\Delta g$ here, which may be found in~\cite{ref:dssv}. 
As can be seen from Tab.~\ref{tab:hessian}, the integral of $\Delta g$
over the RHIC $x$-region 0.05 to 0.2, $\Delta g^{\mathrm{RHIC}}$, is found to be almost zero,
while Tab.~\ref{tab:mom} shows that extrapolation to all $x$ 
results in the gluon spin contribution $\Delta g^1=-0.084$ at 
$Q^2=10$~GeV$^2$. We stress, however, that this result is not yet
reliable due to the large uncertainty in extrapolation to $x\to 0$. In any case,
there are presently no indications of a sizable contribution of gluon
spins to the proton spin. This is in line with recent theoretical expectations
obtained within an effective low-energy theory of broken scale invariance
of QCD~\cite{lkt}. Recent bag model estimates also point to relatively
modest (but positive) values~\cite{dgbag}. Very large values of the 
integral of the spin-dependent gluon distribution, $\Delta g^1\sim 1.5$ or 
so at $Q^2=1$~GeV$^2$, as predicted based on considerations of the QCD axial 
anomaly~\cite{anomaly}, become increasingly disfavored, unless $\Delta g$ 
would show a steep rise at small $x$. 
Future data from RHIC for spin 
asymmetries in forward production of correlated hadron or jet pairs, 
and from running at 500~GeV c.m.s. energy, are expected to shed 
light on $\Delta g$ at lower momentum fractions~\cite{spinplan}.
Again, also 
a polarized electron-ion collider \cite{raju} would be ideally suited to
address this important question and to quantify the amount of gluon polarization 
at small $x$ from measurements of scaling violations of the structure function $g_1$. 
Other promising channels are, for instance, the polarized photoproduction of single-inclusive
hadrons \cite{ref:photo-hadron} or jets \cite{ref:photo-jets}.

We have shown the comparison to some of the RHIC data 
in Fig.~\ref{fig:rhic-all-hessian} (see also Ref.~\cite{ref:dssv}).
A way to access $\Delta g$ in lepton-nucleon scattering
is to measure final states that dominantly select the photon-gluon fusion process,
heavy-flavor production, $\ell p\to h X $, and $\ell p\to h^+ h^- X $, where the
hadrons have large transverse momentum. Figure~\ref{fig:dgdis} 
shows the corresponding results for the extracted $\Delta g/g$  
from SMC, HERMES, and COMPASS \cite{ref:hermes,ref:smc,ref:compass-2had,ref:compass-charm}, 
which have not yet been included in our global analysis. 
We also show in the figure our result, for two representative $Q^2$ scales. 
It should be noted that this comparison is not quite consistent,
as the extraction of $\Delta g/g$ by the experiments was performed
at LO level based on Monte-Carlo generators. Nonetheless, 
a small $\Delta g$ at $x\simeq 0.08-0.2$ as found in our analysis
is also well consistent with the data from lepton-nucleon scattering.
We expect that the data for the measured spin asymmetries will be
included in our global analysis in the future, after the NLO framework
for them has been fully developed and been compared to data for
the corresponding spin-averaged cross sections. 

%%%%%%%%%%%%%%%%%%%%%%%%%%%%%%%%%%%%%%%%%%%%%
\subsection{Exploring the fit parameter space}
%%%%%%%%%%%%%%%%%%%%%%%%%%%%%%%%%%%%%%%%%%%%%
%
In this Section we briefly present a few more details of 
the behavior of our total $\chi^2$ near its minimum, which 
has ramifications, in particular, for the use of the Hessian 
matrix method for estimating uncertainties. As we noted before,
an advantage of the Hessian technique is that it allows to 
produce sets of ``eigenvector PDFs''~\cite{cteq2}, which in 
turn can be straightforwardly used in computations of other
observables, in order to estimate their PDF uncertainty
based on Eq.~(\ref{eq:obserror-hessian}). 
For this, however, it is very important to know the range
of validity of the method, i.e., to which degree $\chi^2$
is parabolic around its minimum.

%%%%%%%%%%%%%%%%%
% FIGURE 11
%%%%%%%%%%%%%%%%%
\begin{figure}[!ht]
\begin{center}
%\vspace*{-0.6cm}
\epsfig{figure=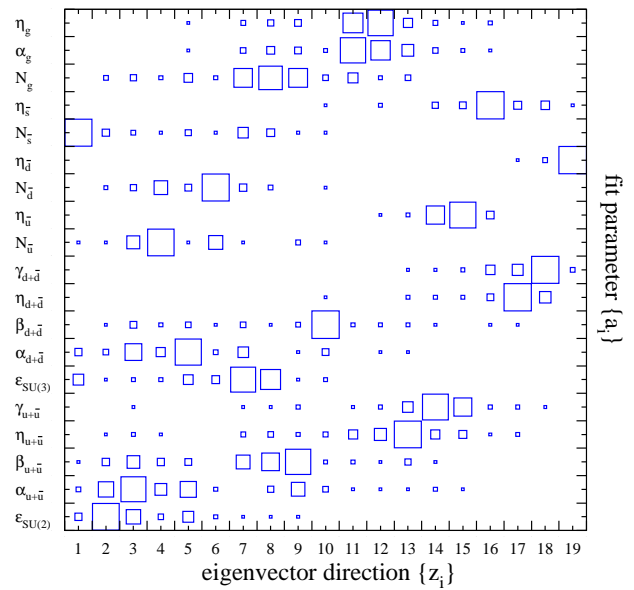,width=0.45\textwidth}
\end{center}
\vspace*{-0.5cm}
\caption{Correlations between the fit parameters $\{a_i\}$ 
and the eigenvector directions $\{z_i\}$. The larger the 
box size the larger the overlap, see text.
\label{fig:boxplot}}
%\vspace*{-0.5cm}
\end{figure} 
 
As described in Sec.~\ref{laghes}, the first step in the Hessian
method is to transform the fit parameters $\{a_i\}$ to a new
set $\{z_i\}$ such that surfaces of constant $\chi^2$ turn 
into hyper-spheres in $\{z_i\}$ space, 
see Eqs.~(\ref{eq:zi-def}), (\ref{eq:deltachi2z})~\cite{cteq2}. 
Figure~\ref{fig:boxplot} shows the overlap of each of 
the original fit parameters $\{a_i\}$ with the eigenvector 
directions $\{z_i\}$; the larger the box size the larger the contribution
of a certain eigenvector direction to a fit parameter $a_i$.
The ${z_i}$ are ordered in terms of the eigenvalues of the Hessian matrix:
$z_1$ corresponds to the largest eigenvalue, i.e., a direction
in parameter space where $\chi^2$ changes rapidly, whereas
$z_{19}$ is only very weakly constrained by data.
One can see that in many cases there is a fairly strong correlation between a 
given original fit parameter $a_i$ and a single eigenvector direction $z_i$.
The parameters which appear to be constrained best by current data are
the normalizations of the sea quarks, $N_{\bar{u}},\,N_{\bar{d}},$ and $N_{\bar{s}}$,
$\varepsilon_{\mathrm{SU}(2,3)}$ controlling the breaking of SU(2,3) symmetry,
and $\alpha_{u+\bar{u}}$, $\alpha_{d+\bar{d}}$ related to the small $x$ behavior
of $\Delta u + \Delta \bar{u}$, $\Delta d + \Delta \bar{d}$.
Parameters determining the gluon distribution, $N_g$, $\alpha_g$, 
and $\eta_g$ are less well constrained and mainly correlated with eigenvector
directions $z_{7}$ to $z_{12}$. $\eta_{\bar{d}}$, $\gamma_{d+\bar{d}}$,
$\eta_{d+\bar{d}}$, and $\eta_{\bar{s}}$ receive contributions from 
eigenvector directions which are only weakly constrained by data.
As we shall see below, this general picture agrees rather well with results
for the $\chi^2$ profiles for each fit parameter
obtained with the Lagrange multiplier method.

\begin{figure}[!ht]
\begin{center}
\vspace*{0.5cm}
\epsfig{figure=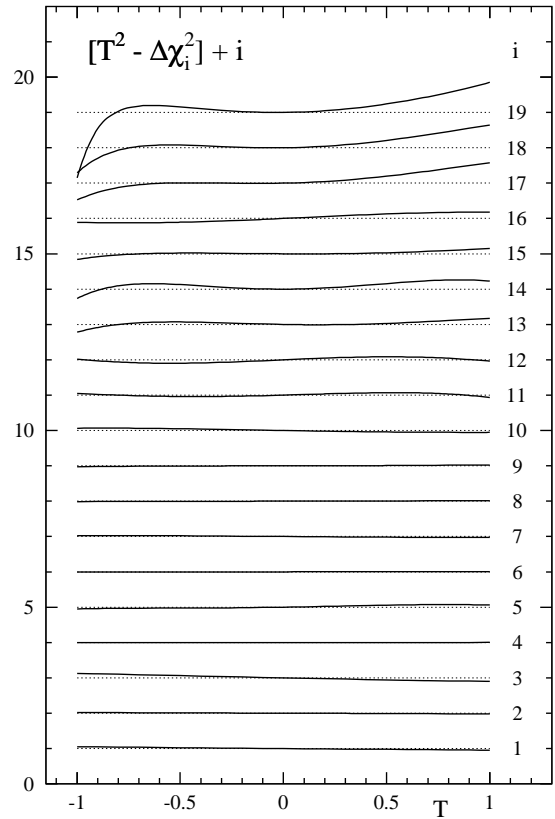,width=0.4\textwidth}
\end{center}
\vspace*{-0.4cm}
\caption{Deviations from the expected parabolic behavior 
$\Delta \chi^2=T^2$ for the eigenvector directions $\{z_i\}$,
see text.
Note that for better separation of the curves we have added an 
off-set $i$ for each parameter $z_i$.
\label{fig:parabolicity}}
%\vspace*{-0.5cm}
\end{figure}

In Figure \ref{fig:parabolicity} we investigate the behavior of $\chi^2$ 
around its minimum, making use of the transformed parameters $\{z_i\}$. 
We vary one of the parameters $z_i$ at a time, keeping all others fixed. 
Of course, since each $z_i$ has in principle overlap
with all fit parameters $\{a_i\}$, the latter all vary in this procedure.
The variation is done in such a way that a given change of $\Delta\chi^2=T$ 
is produced.
For truly quadratic behavior near the minimum, as is the underlying 
assumption in the Hessian approach, the quantity $T^2-\Delta\chi_i^2$,
where $\Delta\chi_i^2$ is the change in $\chi^2$ contributed by
the parameter $z_i$ that is varied, is trivially zero. 
This can be compared to the actual dependence of $\chi^2$ on the varied 
parameter, making no use of the quadratic expansion in~(\ref{eq:hij}). 
Any deviation of $T^2-\Delta\chi_i^2$ from zero will signal a
departure from the quadratic behavior near the minimum. 
One can see from the figure that a choice $\Delta\chi^2=1$ works reasonably
well overall, in the sense that overall only fairly small deviations from 
zero occur. This implies that the Hessian matrix method is reliable
for $\Delta\chi^2=1$ and our eigenvector sets $S_k^{\pm}$ 
will produce faithful uncertainty estimates. 
Some eigenvector directions starting from $z_{12}$ and higher
do show a certain departure from the ideal behavior even for 
$\Delta\chi^2<1$. This is most pronounced for $z_{17}$ to $z_{19}$
which are the least constrained parameters. 
In general we have found that the Hessian method breaks down rapidly 
once one goes beyond $\Delta\chi^2=1$. Therefore we cannot provide
eigenvector sets $S_k^{\pm}$ corresponding to the more conservative 
error estimate $\Delta \chi^2/\chi^2=2\%$ preferred in \cite{ref:dssv}.

%%%%%%%%%%%%%%%%%
% FIGURE 12
%%%%%%%%%%%%%%%%%

\begin{figure*}[!ht]
\begin{center}
\vspace*{-0.6cm}
\hspace*{-1cm}
\epsfig{figure=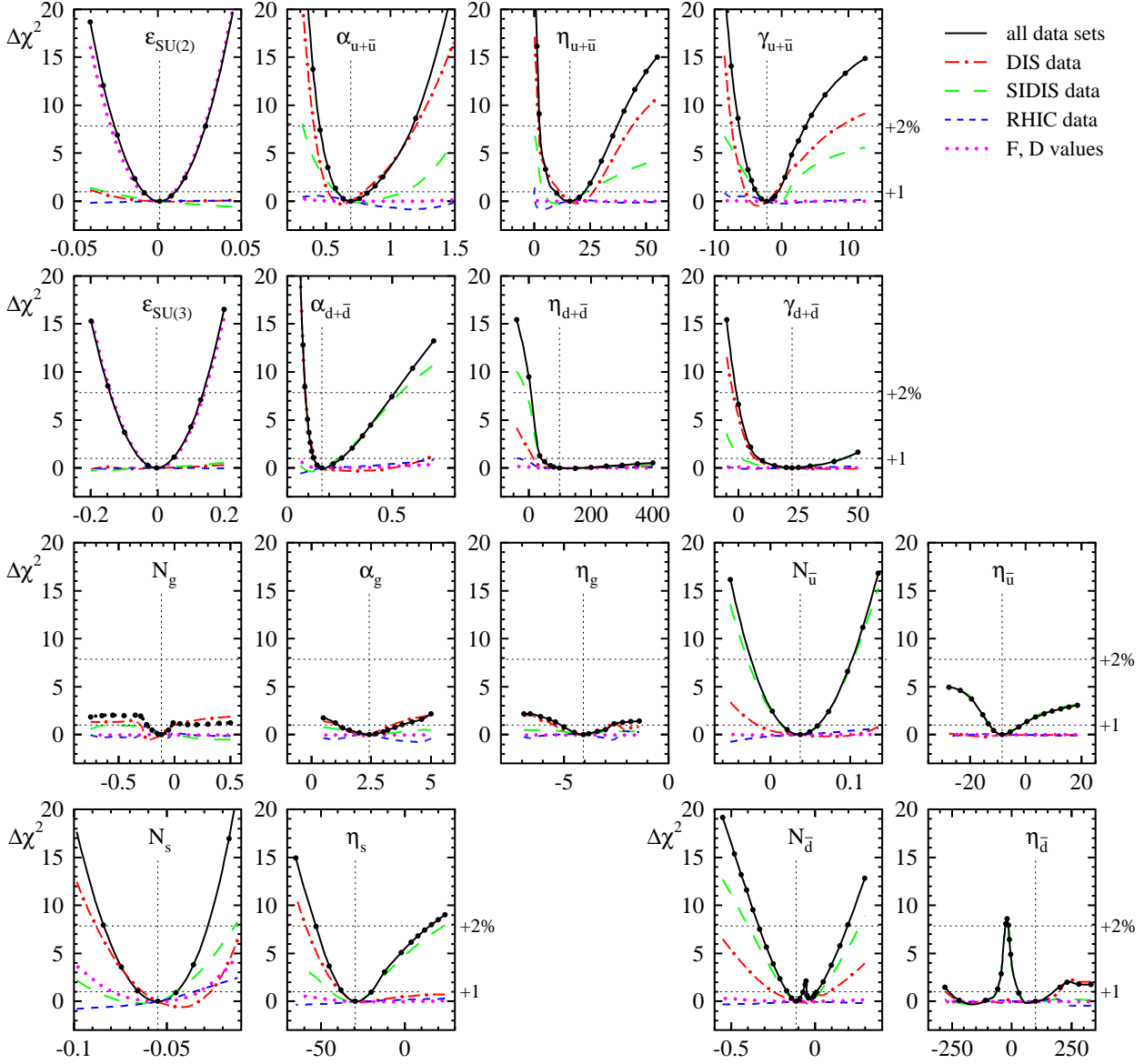,width=1\textwidth}
\end{center}
\vspace*{-0.7cm}
\caption{The $\chi^2$ profiles obtained with the Lagrange multiplier approach for
the parameters $\{a_i\}$ of the fit. The solid lines represent the increase in $\chi^2$ above
the value obtained in the best fit. Dashed-dotted lines, long dashes, dots, 
and short dashed lines represent the partial contributions coming from 
inclusive, semi-inclusive, RHIC and baryon-decay
data, respectively. The parameters $\beta_i$
have been fixed here and are hence not included.
\label{fig:par4}}
\vspace*{-0.5cm}
\end{figure*}

Figure~\ref{fig:par4} shows 
the $\chi^2$ profiles including the individual contributions
from the DIS, SIDIS, and RHIC $pp$ data sets and from the $F$, $D$ values
for the fit parameters $\{a_i\}$, obtained with the Lagrange multiplier approach. 
Clearly, while for some of the parameters the profiles are smooth and 
parabolic as expected in the simplest approach, for others they are not, 
showing not only non-parabolic behavior but variously asymmetric shapes, 
multiple minima or almost flat regions. It is worth pointing out that these 
behaviors are not related to a lack of flexibility of the input
parameterizations, but to features of the data itself. For example, the 
double minima observed for $N_{\overline{d}}$ and $\eta_{\overline{d}}$ 
are associated with two possible ``best-fit solutions'' to the pion sidis 
asymmetries, which show strong fluctuations.

In most cases, the behavior is still reasonably quadratic within
$\Delta \chi^2<1$, however, which further justifies the 
applicability of the Hessian method for $\Delta \chi^2=1$. 
Beyond that, simple extrapolation based
on an assumed quadratic behavior may give misleading results. 
We recall that the central values for the parameters can be 
found in Tab.~\ref{tab:para}.
%
%%%%%%%%%%%%%%%%%
% FIGURE 13
%%%%%%%%%%%%%%%%%

Reducing the number of parameters of the fit would improve the constraints 
on the remaining ones, however at the expense of reducing the
quality of the fit. The resulting constraint in that case would be strongly 
dependent on the functional form assumed for the PDFs. In this sense, the robust
error analysis based on Lagrange multipliers allows to use more flexible functional 
forms.

%%%%%%%%%%%%%%%%%%%%%%%%%%%%%%%%%%%%%%%%%%%%%%%%%%%%%%%%%%%%%%%%%%%%%%%%%%%%%%%
\section{Applications of our uncertainty estimates \label{sec:dijets}}
%%%%%%%%%%%%%%%%%%%%%%%%%%%%%%%%%%%%%%%%%%%%%%%%%%%%%%%%%%%%%%%%%%%%%%%%%%%%%%%
%
The 38 eigenvector PDF sets $S_k^{\pm}$ that we have constructed for the Hessian matrix
method are ideally suited for estimating the PDF uncertainty of
other observables. In this way, one can for example gauge the 
accuracy that future additional measurements will need to have
in order to have a significant impact on our knowledge of the PDFs.
In this Section, we will present a few examples for the 
case of RHIC. In view of the fact that RHIC has just completed its
first physics run at $\sqrt{S}=500$~GeV, we will focus on predictions 
for this c.m.s.\ energy. 

\begin{figure}[!ht]
\begin{center}
\epsfig{figure=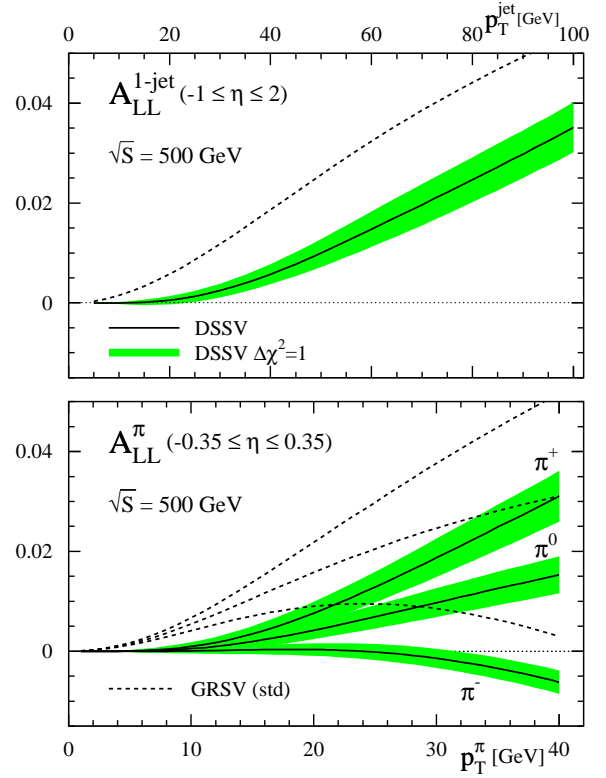,width=0.5\textwidth}
\end{center}
\vspace*{-0.5cm}
\caption{Double-spin asymmetries $A_{LL}$ for jet and pion
production at RHIC at $\sqrt{S}=500$~GeV as functions of the 
transverse momentum $p_T$ of the produced final state. We show 
the results for the best-fit parton distributions from our global 
analysis, along with the uncertainties estimated using the 
Hessian method, allowing changes of one unit in 
$\chi^2$. We also show the results for the ``standard'' scenario
of~\cite{grsv} (dashed lines; in the lower plot the result for 
$\pi^+$ ($\pi^-$) is given by the top (bottom) curve).
We have used the CTEQ6M unpolarized parton 
distributions~\cite{cteq6} for the calculation of the denominator
of the asymmetry. For the pions, we have assumed pseudo-rapidity 
coverage of $|\eta|<0.35$, and for the jets of $-1\leq\eta\leq 2$.
\label{fig:pionsjets500}}
\end{figure}

Figure~\ref{fig:pionsjets500} 
shows the NLO double-spin asymmetries $A_{LL}$ defined in Eq.~(\ref{eq:alldef})
for $pp\to hX$ ($h=\pi^0,\pi^{\pm}$) 
and $pp\to {\mathrm{jet}}X$, for our central DSSV fit (solid lines),
including the Hessian uncertainty bands for $\Delta\chi^2=1$
using Eq.~(\ref{eq:obserror-hessian}).
One can see that the asymmetry for $\pi^0$ remains very small until about
$p_T\sim 20$~GeV, as could be expected from a simple scaling of the 
asymmetry $A_{LL}$ at $\sqrt{S}=200$~GeV shown in Fig.~(\ref{fig:rhic-all-hessian})
with $x_T\equiv 2 p_T/\sqrt{S}$. It then rapidly increases. 
The asymmetry for negatively charged pions 
remains small for all transverse momenta and in fact turns 
slightly negative at high $p_T$. In contrast, $A_{LL}^{\pi^+}$ is 
higher than $A_{LL}^{\pi^0}$, reaching about $3\%$ at the highest
$p_T$ shown. The behavior of the various pion asymmetries is closely
tied to that of the polarized gluon distribution: at high $p_T$,
relatively large values of $x$ are relevant, where our $\Delta g$
is positive, and quark-gluon scattering dominates. An important
contribution to the spin-dependent cross section thus involves the 
combination $(\Delta u \otimes D_u^{\pi}+\Delta \bar{u} \otimes 
D_{\bar{u}}^{\pi}+ \Delta d \otimes D_d^{\pi}+\Delta \bar{d} \otimes 
D_{\bar{d}}^{\pi})\otimes \Delta g$ of parton 
distributions and fragmentation functions. 
For $\pi^+$ production, the $u$ quark and 
$\bar{d}$ anti-quark contributions are expected to dominate,
as these are valence quarks in a $\pi^+$. The combination 
$\Delta u+\Delta \bar{d}$ is positive as Fig.~\ref{fig:bands} shows. 
The large negative contribution associated with $\Delta d$ is suppressed 
here. For $\pi^0$ production, the participating fragmentation
functions are all equal, and one probes the sum of up and down quark
and anti-quark distributions, which is positive but smaller than  
$\Delta u+\Delta \bar{d}$. Finally, for $\pi^-$ production, 
the main contribution involves $\Delta d+\Delta \bar{u}$, which
explains the downturn of $A_{LL}^{\pi^-}$ to negative values
at high $p_T$. Clearly, the three pion asymmetries are also sensitive to 
the sign of $\Delta g$. We note that preliminary 
results for $A_{LL}^{\pi^{\pm}}$
at $\sqrt{S}=200$~GeV have recently been reported from 
RHIC~\cite{rhic:charged}. 

Similar features as for the $\pi^0$ asymmetry are observed for jets. 
Very roughly, one finds that $A_{LL}^{\mathrm{jet}}(p_T)
\approx A_{LL}^{\pi^0}(k p_T)$, where $k\approx 0.5$ or so, 
corresponding to the fact that on average only the fraction $k$ of 
the total jet momentum is taken by an observed $\pi^0$. This implies
that, at a given $p_T$, the jet spin asymmetry is smaller than that
for $\pi^0$.

%%%%%%%%%%%%%%%%%
% FIGURE 14
%%%%%%%%%%%%%%%%%

We have seen in the previous Section that the SIDIS data have
given some first insights into the flavor structure of the 
polarized sea distributions of the nucleon. On the other hand,
the uncertainties in SIDIS are still quite large, and it 
is in particular difficult to quantify the systematic 
uncertainty of the results related to the fragmentation 
mechanism at the relatively modest energies available so far. 
Complementary and clean information on $\Delta u,\,\Delta \bar{u},\,
\Delta d$, and $\Delta \bar{d}$ will come from 
$pp\to W^{\pm}X$ at RHIC, 
where one will exploit the maximally parity-violating couplings of 
produced $W$ bosons to left-handed quarks and right-handed 
anti-quarks~\cite{spinplan,soffer1}.
The high scale set by the $W$ boson mass makes it possible to 
extract quark and anti-quark polarizations from inclusive lepton 
single-spin asymmetries in $W$ boson production with minimal theoretical 
uncertainties, as higher order and sub-leading terms in the perturbative 
QCD expansion are suppressed~\cite{kamal,nadolsky,asmita,ddfwv}. 

%%%%%%%%%%%%%%%%%
% FIGURE 15
%%%%%%%%%%%%%%%%%
\begin{figure}[!ht]
\begin{center}
%\vspace*{5cm}
\epsfig{figure=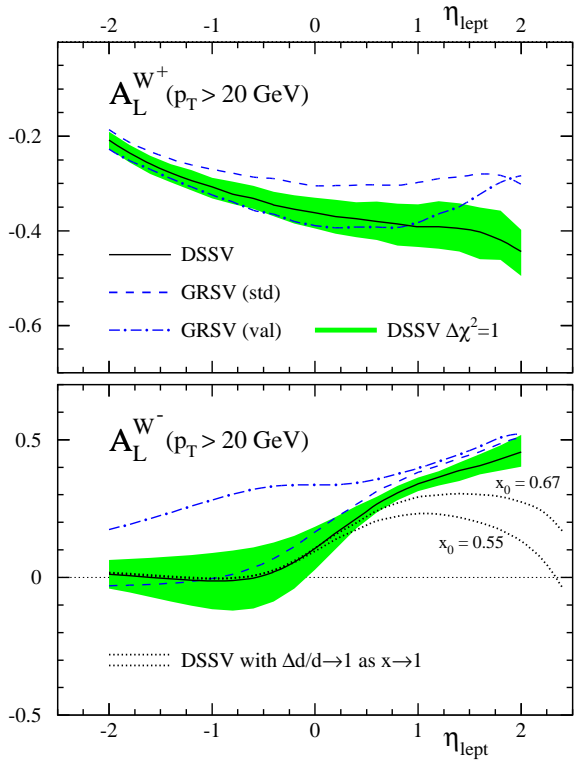,width=0.5\textwidth}
\end{center}
\vspace*{-0.7cm}
\caption{Single-longitudinal spin asymmetries for charged-lepton 
production at RHIC through production and decay of $W$ bosons. The 
bands correspond to our uncertainty estimates based on the Hessian
$\Delta\chi^2=1$ eigenvector PDFs. We also show in the figure 
the spin asymmetries obtained for the ``standard'' and ``valence'' 
scenarios of~\cite{grsv}. For the case of $W^-$, we also show
the results of two fits for which the ratio $\Delta d(x)/d(x)$
is forced to turn to $+1$ as $x\to 1$, see text.
\label{fig:W500}}
%\vspace*{0.5cm}
\end{figure}

As a further application of our Hessian uncertainty PDFs, we
show in Fig.~\ref{fig:W500} the single-longitudinal spin
asymmetries,
\begin{equation}
A_L\equiv\frac{\sigma^{+}-\sigma^{-}}{\sigma^{+}+\sigma^{-}} \, ,
\label{eq:aldef}
\end{equation}
for the processes $\vec{p}p\to \ell^{\pm} X$, where the arrow denotes
a longitudinally polarized proton and $\ell=e$ or $\mu$. The charged lepton 
is assumed to have been produced by a leptonic decay of the $W^{\pm}$ 
boson. The asymmetries are shown as functions of the charged lepton's
rapidity $\eta_{\mathrm{lept}}$, with $\eta_{\mathrm{lept}}$ counted
positive in the forward direction of the polarized proton. We have 
integrated over $p_T>20$~GeV, where $p_T$ is the lepton transverse momentum. 
The results shown in the figure are based on a simple LO calculation
of the processes $q\bar{q}'\to W^{\pm}\to \ell^{\pm}\nu$; the NLO
corrections which we should in principle include for consistency
are negligible for this observable~\cite{kamal,nadolsky,asmita,ddfwv}. 

For $W^-$ production, neglecting all partonic processes but the 
dominant $\bar{u}d \to W^-$ one, the spin-dependent cross section
in the numerator of the asymmetry is found to be proportional to the 
combination~\cite{nadolsky}
\begin{equation}
\Delta \bar{u}(x_1)d(x_2)(1-\cos\theta)^2-
\Delta d(x_1) \bar{u}(x_2)(1+\cos\theta)^2\, ,
\label{eq:w-lo}
\end{equation} 
where $\theta$ is the polar angle of the electron in the partonic
c.m.s., with $\theta=0$ in the forward direction of
the polarized parton. At large negative $\eta_{\mathrm{lept}}$, 
one has $x_2\gg x_1$ and $\theta\gg\pi/2$. In this case, the 
first term in Eq.~(\ref{eq:w-lo}) strongly dominates, since 
the combination of parton distributions, $\Delta\bar{u}(x_1)
d(x_2)$, and the angular factor, $(1-\cos\theta)^2$, each dominate 
over their counterpart in the second term. Since the denominator
of $A_L$ is proportional to $\bar{u}(x_1)d(x_2)(1-\cos\theta)^2+
d(x_1) \bar{u}(x_2)(1+\cos\theta)^2$, the asymmetry provides a
clean probe of $\Delta\bar{u}(x_1)/\bar{u}(x_1)$ at medium
values of $x_1$. Indeed, the Hessian uncertainty band for 
$\Delta\bar{u}$ shown in Fig.~(\ref{fig:bands}) is directly 
reflected in the band we show in Fig.~\ref{fig:W500}. We also show
in the figure the spin asymmetries obtained for the ``standard''
and ``valence'' scenarios of~\cite{grsv}. The latter has a large 
and positive $\Delta\bar{u}$ distribution at the relevant $x\sim 0.1$,
which clearly shows in the asymmetry. By similar 
reasoning, at forward rapidity $\eta_{\mathrm{lept}}\gg 0$ the 
second term in Eq.~(\ref{eq:w-lo}) dominates, giving access
to $-\Delta d(x_1)/d(x_1)$ at relatively high $x_1$. We have
discussed in Subsec.~\ref{sec:extracted} that there is interest
in the question if the polarized down-quark distribution turns 
positive in the large-$x$ region, for which there are currently
no indications, see Fig.~(\ref{fig:largex}). As Fig.~\ref{fig:W500}
shows, the asymmetry for $W^-$ production becomes large and
positive at high $\eta_{\mathrm{lept}}$, which precisely
reflects the fact that $\Delta d(x)$ remains negative at high $x$
in our DSSV fit. It is interesting to investigate how the asymmetry
might look if $\Delta d/d$ were to turn to $+1$ as $x\to 1$.
In order to do this, we have produced two fits where $\Delta d/d$
is forced to have this behavior. The two fits are characterized
by the value $x_0$ where $\Delta d(x,M_W^2)$ changes 
sign from negative to positive values. We have chosen $x_0=0.67$
and $x_0=0.55$. The $\chi^2$ values for these two fits are
of course significantly worse than for our DSSV best fit, by
about four units for $x_0=0.67$ and about 25 units for $x_0=0.55$.
The results for the two fits are shown by the dotted lines 
in Fig.~\ref{fig:W500}. It should be well possible at RHIC to 
measure the asymmetry for values $\eta_{\mathrm{lept}}$ out to
$\gtrsim 2$~\cite{spinplan}. We note that the 
behavior of $\Delta d/d$ at high $x$ will also be further addressed 
by experiments at Jefferson Lab after the 12~GeV upgrade~\cite{ref:bruell}. 

For $W^+$ production, one has the following structure of the 
spin-dependent cross section~\cite{nadolsky}:
\begin{equation}
\Delta \bar{d}(x_1)u(x_2)(1+\cos\theta)^2-
\Delta u(x_1) \bar{d}(x_2)(1-\cos\theta)^2\, .
\label{eq:w+lo}
\end{equation} 
Here the distinction of the two contributions by considering 
large negative or positive lepton rapidities is less clear-cut than
in the case of $W^-$. For example, at negative $\eta_{\mathrm{lept}}$
the partonic combination $\bar{d}(x_1)u(x_2)$ will dominate, but
at the same time $\theta\gg\pi/2$ so that the angular 
factor $(1+\cos\theta)^2$ is small. Likewise, at positive 
$\eta_{\mathrm{lept}}$ the dominant partonic combination 
$\Delta u(x_1) \bar{d}(x_2)$ is suppressed by the angular factor.
So both terms in Eq.~(\ref{eq:w+lo}) will compete essentially 
for all $\eta_{\mathrm{lept}}$ of interest. This is reflected 
in the behavior of the calculated spin asymmetry $A_L^{W^+}$ 
shown in Fig.~\ref{fig:W500}, which does not show as clear features
as the one for $W^-$ bosons. Nonetheless, the $W^+$ measurements
at RHIC will of course still be of great value. In fact, our global 
analysis technique is precisely suited for extracting information on the 
polarized PDFs even if there is no single dominant partonic 
subprocess.

%%%%%%%%%%%%%%%%%%%%%%%%%
\section{Conclusions}
%%%%%%%%%%%%%%%%%%%%%%%%%
%
We have presented details of a recent study of the helicity parton distribution
functions of the nucleon, which used experimental information available from 
inclusive and semi-inclusive polarized deep-inelastic lepton-nucleon 
scattering and from polarized proton-proton scattering at RHIC.
The data sets were used jointly in a next-to-leading order
global QCD analysis, which allows to extract the set of parton distributions
that provides the optimal overall description of the data, along with
estimates of its uncertainties. We have presented techniques and
computational methods that speed up the next-to-leading order 
calculations for $pp$ scattering to the level required in practice 
for a global analysis. Our technique is formulated in Mellin moment space. 
A key feature is that the computationally most challenging parts
are done only once, prior to the fit. Use of a Monte-Carlo sampling 
method allows us to perform this one-time calculation very efficiently.

Our extracted parton distributions show particularly interesting features
in the sea quark and gluon sector. We find evidence for a mostly positive 
$\Delta \bar{u}$ and a negative $\Delta \bar{d}$ distribution, so that
$\Delta \bar{u}-\Delta \bar{d}$ is positive. This behavior has been 
predicted by a number of models of nucleon structure. The polarized
strange quark distribution $\Delta s$ comes out slightly positive at 
medium $x$, which is driven by the semi-inclusive kaon DIS data and 
could be subject to rather large systematic uncertainties. $\Delta s$
turns negative at $x\lesssim 0.02$ as a result of constraints from
SU(3) symmetry, which have a relatively small nominal error. If true,
this means that $\Delta s$ acquires its large negative integral 
essentially completely from the small-$x$ region. As a further
consequence, quark and anti-quark spins combined contribute about
a fourth to a third of the proton's spin, with the lower value
arising if strange quarks and anti-quarks are indeed strongly
negatively polarized at low $x$. Finally, we have found that
the gluon helicity distribution $\Delta g(x,Q^2)$ is small in the
region of momentum fraction accessed directly so far by RHIC,
with likely a node and an almost vanishing integral over that region. 
Reliable statements about the full gluon spin contribution to the 
proton spin are presently not yet possible. 

We have performed uncertainty estimates for our polarized parton 
distributions, using both the Lagrange multiplier technique and
the improved Hessian approach. To obtain these, a large number
of additional fits are necessary, for which the 
computational techniques we
have developed are particularly important. We find that both 
approaches yield consistent results for moderate departures 
from the best fit, typically $\Delta \chi^2=1$. For larger
$\Delta \chi^2$, significant differences develop as a result
of departures from parabolic behavior of $\chi^2$ around its
minimum. This implies that the Hessian matrix method becomes 
unreliable. We have produced a set of 38 ``eigenvector'' parton
distributions for the Hessian method with $\Delta \chi^2=1$ \cite{ref:webpage},
which may be used to estimate the uncertainty of any observable 
that depends on the distributions. We stress, however, that we presently 
prefer a more conservative choice of $\Delta \chi^2/\chi^2=2\%$
as a tolerance criterion for acceptable parton distributions.
Unfortunately, the behavior of $\chi^2$ around its minimum
does not warrant use of the Hessian method for producing 
eigenvector parton distributions in this case.

We have used the $\Delta \chi^2=1$ eigenvector distributions to 
obtain predictions for spin asymmetries for high transverse momentum 
pion and jet production in polarized proton-proton collisions at 
500~GeV center-of-mass energy at BNL-RHIC, as well as for $W$ 
boson production. The former would give information on $\Delta g$ 
at lower $x$, while the latter would provide a clean new probe 
of the polarized quark and anti-quark distributions, which 
is important in view of the uncertainties inherent in semi-inclusive 
DIS. Our results indicate that there is significant potential for 
RHIC to provide further important insights into nucleon helicity 
structure. It will be straightforward to include all the 
forthcoming data in the global analysis.

%%%%%%%%%%%%%%%%
\acknowledgments
%%%%%%%%%%%%%%%%
We thank E.C.\ Aschenauer, H.\ Avakian, A.\ Bazilevsky, K.\ Boyle,
A.\ Deshpande, S.\ Kuhn, Z. Meziani, D.\ Stamenov, B.\ Surrow, S.\ Taneja, 
R.\ Thorne, C.\ Weiss, and F.\ Yuan for communications and discussions.          
W.V.\ is grateful to the U.S.\ Department of Energy 
(contract number DE-AC02-98CH10886) for
providing the facilities essential for the completion of his work.
This work was supported in part by LDRD project 08-004 of 
Brookhaven National Laboratory. 
M.S.\ acknowledges partial support of the Initiative and
Networking Fund of the Helmholtz Association, contract HA-101
("Physics at the Terascale") and the ``Bundesministerium f\"ur
Bildung und Forschung'' (BMBF), Germany.
This work was partially supported by CONICET, ANPCyT and UBACyT.
D.\ de F.'s work was supported by a fellowship by the John Simon Guggenheim 
Memorial Foundation.

%%%%%%%%%%%%%%%%%%%%%%%%%%%

\end{document}